\documentclass[iop]{emulateapj}   		

\usepackage{subfigure}					
\usepackage{graphicx}					
\usepackage{amssymb}					
\usepackage{mathtools}					
\usepackage{natbib}						

\usepackage{color}						
\usepackage{amsmath}
\usepackage{bm}						

\begin{document}

\title{Searching For Fossil Evidence of AGN Feedback in \textit{WISE}-Selected Stripe-82 Galaxies By Measuring the Thermal Sunyaev-Zel'dovich Effect With the Atacama Cosmology Telescope}
\author{Alexander Spacek, Evan Scannapieco, Seth Cohen, Bhavin Joshi, Philip Mauskopf}						
\affil{School of Earth and Space Exploration, Arizona State University, P.O. Box 876004, Tempe, AZ 85287, USA}


\begin{abstract}

We directly measure the thermal energy of the gas surrounding galaxies through the thermal Sunyaev-Zel'dovich (tSZ) effect. We perform a stacking analysis of microwave background images from the Atacama Cosmology Telescope, around 1179 massive quiescent elliptical galaxies at $0.5 \leq z \leq 1.0$ (``low-$z$'') and 3274 galaxies at $1.0 \leq z \leq 1.5$ (``high-$z$''), selected using data from the \textit{Wide-Field Infrared Survey Explorer} All-Sky Survey and the Sloan Digital Sky Survey (SDSS) within the SDSS Stripe-82 field. The gas surrounding these galaxies is expected to contain energy from past episodes of active galactic nucleus (AGN) feedback, and after using modeling to subtract undetected contaminants, we detect a tSZ signal at a significance of $0.9\sigma$ for our low-$z$ galaxies and $1.8 \sigma$ for our high-$z$ galaxies. We then include data from the high-frequency \textit{Planck} bands for a subset of 227 low-$z$ galaxies and 529 high-$z$ galaxies and find low-$z$ and high-$z$ tSZ detections of $1.0\sigma$ and $1.5\sigma$, respectively. These results indicate an average thermal heating around these galaxies of $(5.6^{+5.9}_{-5.6}) \times 10^{60}$ erg for our low-$z$ galaxies and $(7.0^{+4.7}_{-4.4}) \times 10^{60}$ erg for our high-$z$ galaxies. Based on simple heating models, these results are consistent with gravitational heating without additional heating due to AGN feedback.

\end{abstract}



\section{Introduction}
\label{sec:intro}

Galaxy formation was long expected to proceed hierarchically, with larger galaxies forming at later times when larger dark matter halos coalesce and gas has longer to cool and condense \citep[e.g.][]{Rees1977,White1991,Richstone1998,Cattaneo1999,Kauffmann2000,Menci2006}. However, an increasing amount of observational evidence suggests recent anti-hierarchical trends in both galaxy and active galactic nucleus (AGN) evolution \citep{Kang2016,RosasGuevara2016,Siudek2016}. For example, beyond $z\approx2$ the typical mass of star-forming galaxies has dropped by more than a factor of $\approx$ 3 \citep{Cowie1996,Brinchmann2004,Kodama2004,Bauer2005,Bundy2005,Feulner2005,Treu2005,Papovich2006,Noeske2007,Cowie2008,Drory2008,Vergani2008}, while the characteristic AGN luminosity has dropped by more than a factor of $\approx$ 10 \citep{Pei1995,Ueda2003,Barger2005,Buchner2015}. To explain these trends, it is widely suggested that galaxy evolution models require additional heating of the circumgalactic medium by energetic AGN feedback \citep{Merloni2004,Scannapieco2004,Scannapieco2005,Bower2006,Neistein2006,Thacker2006,Sijacki2007,Merloni2008,Chen2009,Hirschmann2012,Mocz2013,Hirschmann2014,Lapi2014,Schaye2015,Keller2016}. This typically involves an energetic AGN outflow caused by radiation pressure or jets that blow material out of the galaxy, heating the nearby intergalactic medium enough to suppress further generations of stars and AGNs.

In fact, there is abundant observational evidence of AGN feedback in action in galaxy clusters \citep{Schawinski2007,Rafferty2008,Fabian2012,Farrah2012,Page2012,Teimoorinia2016}. Most notably, the centers of clusters are more likely to contain galaxies that host large radio-loud jets of AGN-driven material \citep{Burns1990,Best2005,McNamara2005}, whose energies are comparable to those needed to stop the gas from cooling \citep[e.g.][]{Simionescu2009}. Furthermore, AGN feedback from the central cD galaxies in clusters increases in proportion to the cooling luminosity, as expected in an operational feedback loop \citep[e.g.][]{Birzan2004,Rafferty2006}.

Direct measurements of AGN feedback in less dense environments are much less common, primarily because of the relatively high redshifts and faint signals involved, although evidence of AGN feedback has also been seen in relatively nearby galaxies \citep{Tombesi2015,Lanz2016,Schlegel2016}. For example, broad absorption-line outflows are observed as blueshifted troughs in the rest-frame spectra of $\approx$20\% of all of quasars \citep{Hewett2003,Ganguly2008,Knigge2008}, but quantifying AGN feedback requires estimating mass-flow and the energy released by these outflows \citep[e.g.][]{Wampler1995,deKool2001,Hamann2001,Feruglio2010,Sturm2011,Veilleux2013}. These quantities can only be computed in cases for which it is possible to estimate the distance to the outflowing material from the central source, which is often highly uncertain. While these measurements have been carried out for a select set of objects \citep[e.g.][]{Chartas2007,Moe2009,Dunn2010,Green2012,Borguet2013,Chamberlain2015}, it is still unclear how these results generalize to AGNs as a whole. Furthermore, it is still an open question as to whether AGN outflows triggered by galaxy interactions in massive, high-redshift galaxies actually quench star formation \citep[e.g.][]{Fontanot2009,Pipino2009,Debuhr2010,Ostriker2010,Faucher2012,Newton2013,Feldmann2015,Bongiorno2016}.

One way to get around the high redshifts and faint signals involved in AGN feedback measurements is to stack measurements of the cosmic microwave background (CMB) radiation. At angular scales smaller than $\approx$5 arcmin, Silk damping washes out the primary CMB anisotropies \citep{Silk1968,PlanckCollaboration2015a}, leaving room for secondary anisotropies such as the Sunyaev-Zel'dovich (SZ) effect, where CMB photons interact with hot, ionized gas \citep{Sunyaev1970,Sunyaev1972}. When the gas has a bulk velocity, CMB photons interacting with electrons in the gas experience a Doppler boost, resulting in frequency-independent fluctuations in the CMB temperature, known as the kinematic Sunyaev-Zel'dovich (kSZ) effect. Although the kSZ effect does not measure the energy of the gas, it can be used to detect the ionized gas profile within dark matter halos, thereby providing information on where hot gas is located around galaxies. This can be useful for understanding how AGN feedback heats up gas and moves it around \citep{Battaglia2010}. Several recent studies have made significant measurements of the kSZ effect in galaxy clusters by stacking CMB observations \citep[e.g.][]{PlanckCollaboration2016,Schaan2016,Soergel2016}.

If the gas is sufficiently heated, inverse Compton scattering coupled with the thermal motions of electrons will shift the CMB photons to higher energies. This thermal Sunyaev-Zel'dovich (tSZ) effect directly depends on the temperature of the free electrons that the CMB radiation passes through, and it has a unique redshift-independent spectral signature that makes it well suited for measuring the gas heated through AGN feedback \citep{Voit1994,Birkinshaw1999,Natarajan1999,Platania2002,Lapi2003,Chatterjee2007,Chatterjee2008,Scannapieco2008,Battaglia2010}.
Individual tSZ distortions per source are very small, however, so a stacking analysis must be performed on many measurements in order to derive a significant signal. 

This method has been used previously by a handful of studies in relation to AGNs and galaxies. \citet{Chatterjee2010} found a tentative detection of quasar feedback using the Sloan Digital Sky Survey (SDSS) and \textit{Wilkinson Microwave Anisotropy Probe}, though the significance of AGN feedback in their measurements is disputed \citep[][]{Ruan2015}. \cite{Hand2011} stacked $>$2300 SDSS-selected ``luminous red galaxies'' in data from the Atacama Cosmology Telescope (ACT) and found a $2.1\sigma-3.8\sigma$ tSZ detection after selecting radio-quiet galaxies and binning them by luminosity. \citet{PlanckCollaboration2013} investigated the relationship between tSZ signal and stellar mass using $\approx260,000$ ``locally brightest galaxies'' with significant results, especially with stellar masses $\gtrsim10^{11} M_\odot$. \cite{Gralla2014} stacked data from the ACT at the positions of a large sample of radio AGNs selected at 1.4 GHz to make a 5$\sigma$ detection of the tSZ effect associated with the haloes that host active AGNs.  \citet{Greco2015} used \textit{Planck} full-mission temperature maps to examine the stacked tSZ signal of 188,042 ``locally brightest galaxies'' selected from the SDSS Data Release 7, finding a significant measurement of the stacked tSZ signal from galaxies with stellar masses above $\approx 2 \times 10^{11} M_\odot$.  \citet{Ruan2015} stacked \textit{Planck} tSZ Compton-$y$ maps centered on the locations of 26,686 spectroscopic quasars identified from SDSS to estimate the mean thermal energies in gas surrounding such $z \approx 1.5$  quasars to be $\approx {10^{62}}$ erg, although the significance of AGN feedback in their measurements has also been disputed \citep{Cen2015}. \citet{Crichton2016} stacked $>$17,000 radio-quiet quasars from the SDSS in ACT data and found $3\sigma$ evidence for the presence of associated thermalized gas and $4\sigma$ evidence for the thermal coupling of quasars to their surrounding medium. \citet{Spacek2016} stacked 937 massive elliptical galaxies using the South Pole Telescope (SPT) and made a $3.6\sigma$ detection of the tSZ effect at a magnitude suggesting an excess of non-gravitational thermal energy, possibly due to AGN feedback. These tSZ AGN feedback measurements continue to be promising, and in this paper we will especially focus on the methods and results from \citet{Spacek2016}.

As evidenced above, quasars are a popular target for measuring AGN feedback due to their brightness and their active feedback processes, but they have drawbacks due to their relative scarcity and the contaminating emission they contain that obscures the tSZ signatures of AGN feedback. In this paper, we follow \citet{Spacek2016} and focus on measuring co-added tSZ distortions in the CMB around massive ($\geq10^{11} M_\odot$) quiescent elliptical galaxies at moderate redshifts ($0.5\leq z \leq 1.5$). To accomplish this, we use data from the \textit{Wide-Field Infrared Survey Explorer} \citep[\textit{WISE};][]{Wright2010}, SDSS \citep[][]{Alam2015}, and ACT \citep{Dunner2013}. These galaxies contain almost no dust and are numerous on the sky, making them well-suited for co-adding in large numbers in order to obtain good constraints on the energy stored in the surrounding gas.

The structure of this paper is as follows: in Section \ref{sec:methods}, we explain our method of measuring the thermal energy around galaxies using the tSZ effect and how that might relate to non-gravitational heating by AGN feedback. In Section \ref{sec:data}, we discuss the data that we use to both select galaxies and make our tSZ measurements. In Section \ref{sec:galaxies}, we outline our galaxy selection process and how we estimate the physical parameters of the galaxies. In Section \ref{sec:filter}, we explain how we filter the ACT images. In Section \ref{sec:stack}, we discuss our stacking procedure and results. In Section \ref{sec:contaminant}, we use a $\chi^2$ analysis to model and remove contaminant signal, and in Section \ref{sec:planck} we do the same but with data from \textit{Planck} included. In Section \ref{sec:discussion}, we summarize our results, discuss the implications for AGN feedback, and provide conclusions.

Throughout this work, we adopt a $\Lambda$CDM cosmological model with parameters \citep[from][]{PlanckCollaboration2015a}, $h=0.68$, $\Omega_0$ = 0.31, $\Omega_\Lambda$ = 0.69, and $\Omega_b = 0.049$, where $h$ is the Hubble constant in units of 100 km s$^{-1}$ Mpc$^{-1}$, and $\Omega_0$, $\Omega_\Lambda$, and $\Omega_b$ are the total matter, vacuum, and baryonic densities, respectively, in units of the critical density. All of our magnitudes are quoted in the AB system \citep[i.e.][]{Oke1983}.



\clearpage
\section{Methods}
\label{sec:methods}

The tSZ effect is the process by which CMB photons gain energy when passing through ionized gas \citep{Sunyaev1970,Sunyaev1972} through inverse Compton scattering with energetic electrons. The resulting CMB anisotropy has a distinctive frequency dependence, which causes a deficit of photons at frequencies below $\nu_{\text{null}} = 217.6 \, \text{GHz}$ and an excess of photons above $\nu_{\text{null}}$. The change in CMB temperature $\Delta T$ as a function of frequency due to the (nonrelativistic) tSZ effect is
\begin{equation} 
\frac{\Delta T}{T_{\text{CMB}}} = y \left( x \frac{e^x + 1}{e^x - 1} - 4 \right),
\label{eq:DeltaT}
\end{equation}
where the dimensionless Compton-$y$ parameter is defined as
\begin{equation}
 y \equiv \int dl \, \sigma_T \frac{n_e k \left( T_e - T_{\rm CMB} \right)}{m_e c^2}, 
 \label{eq:y}
\end{equation}
 where $\sigma_T$ is the Thomson cross section, $k$ is the Boltzmann constant, $m_e$ is the electron mass,
 $c$ is the speed of light, $n_e$ is the electron number density, $T_e$ is the electron temperature, $T_{\text{CMB}}$ is the CMB temperature (we use $T_{\text{CMB}} = 2.725$ K), the integral is performed over the line-of-sight distance $l$, and the dimensionless frequency $x$ is given by $x \equiv \frac{h \nu}{k T_{\text{CMB}}} = \frac{\nu}{56.81 \, \text{GHz}}$, where $h$ is the Planck constant. We can calculate the total excess thermal energy $E_{\rm therm}$ associated with a source by integrating Equation (\ref{eq:y}) over a region of sky around the source, as detailed in \citet{Spacek2016}, and combining the result with Equation (\ref{eq:DeltaT}) to get $E_{\rm therm}$ as a function of $x$ and $\Delta T$. This gives 
\begin{equation}
E_{\rm therm} = \frac{1.1 \times 10^{60} {\rm ergs}}{x \frac{e^x + 1}{e^x - 1} - 4} \, \left(\frac{l_{\rm ang}}{\text{Gpc}}\right)^2 
\frac{\int \Delta T(\bm{\theta}) d\bm{\theta}}{\text{$\mu$K arcmin$^2$}}.
\label{eq:EthrmT}
\end{equation}

To compare the measured energies above with the expectations from models of feedback, we use the simple models of gas heating with and without AGN feedback worked out in \citet{Spacek2016}. For purely gravitational heating, we can assume that the gas collapses and virializes along with an encompassing spherical halo of dark matter. The gas is shock-heated during gravitational infall to a virial temperature $T_{\rm vir}$, and if we approximate the gas as isothermal at this temperature we can estimate its total thermal energy as
\begin{equation}
\begin{split}
E_{\rm therm,gravity} &= \frac{3 k T_{\rm vir}}{2} \frac{\Omega_b}{\Omega_0} \frac{M}{\mu m_p}\\  &=
1.5 \times 10^{60} \, {\rm erg} \, M_{13}^{5/3} (1+z),
\label{eq:Etherm}
\end{split}
\end{equation}
where $m_p$ is the proton mass, $\mu = 0.62$ is the average particle mass in units of $m_p$, and $M_{13}$ is the mass of the halo in units of $10^{13} M_\odot.$
We can convert from halo mass to the stellar mass of the galaxies we will be measuring if we use the observed relation between black hole mass and halo circular velocity $v_c$ 
from \citet{Ferrarese2002}, convert the black hole mass to its corresponding bulge dynamical mass using a factor of 400 \citep{Marconi2003}, use the fact that $v_c = (G M/R_{\rm vir})^{1/2} = 254 \, {\rm km \, s}^{-1} \, M_{13}^{1/3} (1+z)^{1/2},$ where $G$ is the gravitational constant, and take $M_{\rm stellar} \propto v_c^5$. As shown in \cite{Spacek2016}, this gives $M_{\rm stellar} 
 = 2.8_{-1.4}^{+2.4}  \times 10^{10} M_\odot   \,  M_{13}^{5/3} (1+z)^{5/2},$ and
substituting this into Equation (\ref{eq:Etherm}) yields
\begin{equation}
\begin{split}
& E_{\rm therm,gravity}  = \\ & 5.4_{-2.9}^{+5.4} \times 10^{60} \, {\rm erg} \,  \frac{M_{\rm stellar}}{10^{11} M_\odot} (1+z)^{-3/2}.
\label{eq:Egrav}
\end{split}
\end{equation}
This is the total thermal energy expected around a galaxy of stellar mass $M_{\rm stellar}$ ignoring both radiative cooling, which will decrease $E_{\rm therm},$ and AGN feedback, which will increase it.

For heating due to AGN feedback it is difficult to be precise because little is known about the dominant mechanism by which AGN feedback operates, and as a result there are many models, each of which leads to somewhat different signatures in our data. We can, however, try to estimate the overall magnitude of AGN feedback heating by making use of the simple model described in \citet{Scannapieco2004}. This is characterized as the heating of gas by a fraction $\epsilon_{k}$ of the total bolometric luminosity of the AGN, where the black hole shines at the Eddington luminosity ($1.2 \times 10^{38}$ erg ${\rm s}^{-1}$ $M_\odot^{-1}$ ) for a time $0.035 \, t_{\rm dynamical}$, with $t_{\rm dynamical} \equiv R_{\rm vir}/v_c = 2.6 \, {\rm Gyr} \, (1+z)^{-3/2}$, where $R_{\rm vir}$ is the halo virial radius. This gives
\begin{equation}
\begin{split}
& E_{\rm therm, feedback} = \\ & 4.1 \times 10^{60} \, {\rm erg} \,  \epsilon_{k,0.05}  \,  \frac{M_{\rm stellar}}{10^{11} M_\odot} \, (1+z)^{-3/2},
\label{eq:EAGN}
\end{split}
\end{equation}
where $\epsilon_{k,0.05} \equiv \epsilon_{k}/0.05$. In this case, 5\% is a typical, though still very uncertain, efficiency needed to achieve anti-heirarchical galaxy evolution through effective feedback \citep[e.g.][]{Scannapieco2004,Thacker2006,Costa2014}.

It is evident that our simple model of feedback energy falls within the errors of our model for gravitational energy, indicating that the differences between models with and without AGN feedback are subtle. Detailed simulations beyond the scope of this paper will be needed to make precise predictions regarding particular AGN feedback models. Even so, our models above are roughly consistent with more sophisticated models \citep[e.g.][]{Thacker2006,Chatterjee2008}, and we will therefore use them to provide a general context for our results.


\section{Data}
\label{sec:data}

In order to select a large number of galaxies for our stacking analysis, we wanted to use a large region of the sky that was covered with a wide wavelength range of telescope surveys and included microwave data for our tSZ measurements. We therefore chose the SDSS Stripe-82 region, which is covered by many surveys, including ultraviolet, visible, and infrared data from the SDSS, infrared data from the \textit{WISE} All-Sky Data Release, and microwave data from the ACT. We also used the extensive pre-existing source catalogs corresponding to the SDSS and \textit{WISE} data.

Our SDSS data were taken from Data Release 12 (DR12) of the third generation of the Sloan Digital Sky Survey  \citep[SDSS-III;][]{Alam2015}. Since 2000 the SDSS has used a 2.5 m wide-field telescope at Apache Point Observatory in New Mexico to image roughly one-third of the sky (31,637 deg$^2$), and the catalog contains information on over 1 billion objects. The SDSS bands we used are $u$, $g$, $r$, $i$, and $z$, with average wavelengths of 355.1, 468.6, 616.5, 748.1, and 893.1 nm, respectively, and an average point spread function (PSF) width of 1.4 arcsec in $r$-band.\footnote{http://classic.sdss.org/dr7/} Stripe-82 is a 2.5$^\circ$ wide stripe along the celestial equator that was imaged multiple times, resulting in deeper SDSS data than the main survey. The general Stripe-82 region runs from $-1.25^\circ$ to $1.25^\circ$ decl. and $-65^\circ$ to $60^\circ$ R.A., with an area of $\approx$312 deg$^2$.\footnote{http://classic.sdss.org/dr7/coverage/sndr7.html}

The \textit{WISE} All-Sky Data Release contains data from the full \textit{WISE} mission in 2010 using the 0.4 m space telescope \citep{Wright2010}. The four \textit{WISE} infrared bands are labelled $W1$, $W2$, $W3$, and $W4$, and are centered at 3.4, 4.6, 12, and 22 $\mu$m, respectively, with average PSF full-width-at-half-maximum (FWHM) values of 6.1, 6.4, 6.5, and 12.0 arcsec, respectively. The corresponding source catalog contains over 500 million sources with a signal-to-noise ratio (S/N) $>$5.\footnote{http://wise2.ipac.caltech.edu/docs/release/allsky/}

The ACT is a 6 m telescope on Cerro Toco in Chile which started observing in 2007. It was equipped with the Millimeter Bolometric Array Camera (MBAC), with bands at 148, 220, and 277 GHz \citep{Dunner2013}. The data used in this paper covering the equatorial Stripe-82 region are from ACT seasons 3 and 4 (2009 and 2010) using the 148 and 220 GHz bands.\footnote{http://lambda.gsfc.nasa.gov/product/act/act\_prod\_table.cfm} These have beam FWHM values of $\approx$1.44 and $\approx$1.08 arcmin, respectively. We used the data designated as ``src\_free,'' where flux from point sources has been removed. Since both seasons cover the same region of sky, we are able to average them together to increase our S/N.

The ACT bands at 148 and 220 GHz are ideal for our tSZ measurements because 148 GHz is close to the peak of the undistorted CMB spectrum (160 GHz) and will see a significant decrement, while 220 GHz is very close to a frequency where the tSZ effect has no effect ($\nu_{\text{null}} = 217.6 \, \text{GHz}$). Equation (\ref{eq:DeltaT}) can be rewritten for the ACT bands after integrating over their filter curves and solving for the Compton-$y$ parameter. The filter curves are taken from \citet{Swetz2011}. This gives
\begin{equation} 
y = -0.38\,\frac{\Delta T_{148}}{1 \text{K}} \hspace{1cm} {\rm and}  \hspace{1cm} y = 5.4\,\frac{\Delta T_{220}}{1 \text{K}},
\label{eq:newys}
\end{equation}
where $\Delta T_{148}$ and $\Delta T_{220}$ are the temperature anisotropies at 148 and 220 GHz. We can similarly integrate Equation (\ref{eq:EthrmT}) over the ACT 148 GHz filter curve to get the total thermal energy surrounding a galaxy as a function of the 148 GHz tSZ decrement,
\begin{equation} 
E_{\rm therm} = -1.1\times 10^{60} \text{ergs} \left(\frac{l_{\rm ang}}{\text{Gpc}}\right)^2 \frac{\int \Delta T_{148}(\bm{\theta}) d\bm{\theta}}{\text{$\mu$K arcmin$^2$}}.
\label{eq:newEthrm}
\end{equation}

\begin{figure*}[ht]
\centerline{\includegraphics[height=10cm]{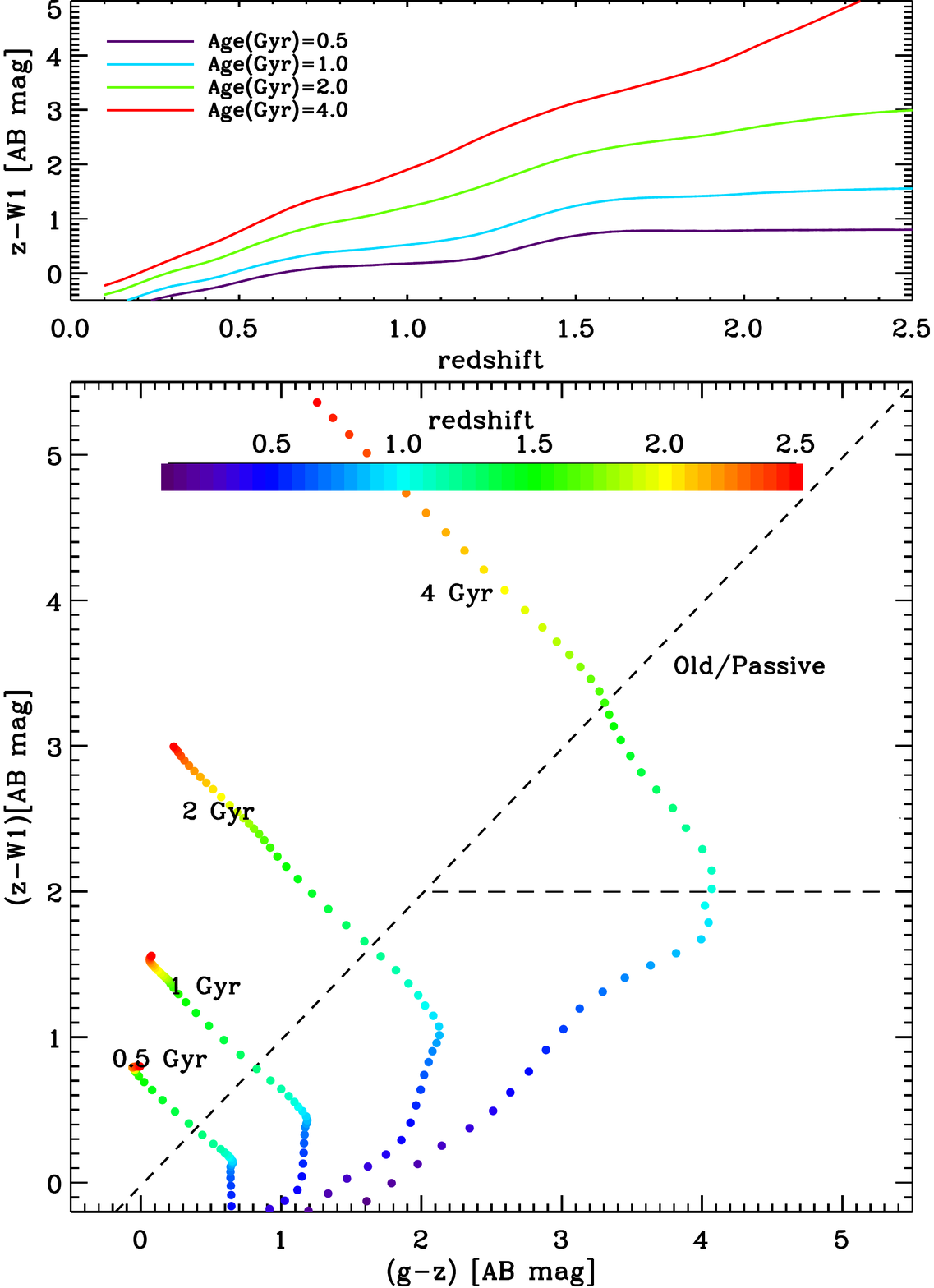}
\includegraphics[height=10cm]{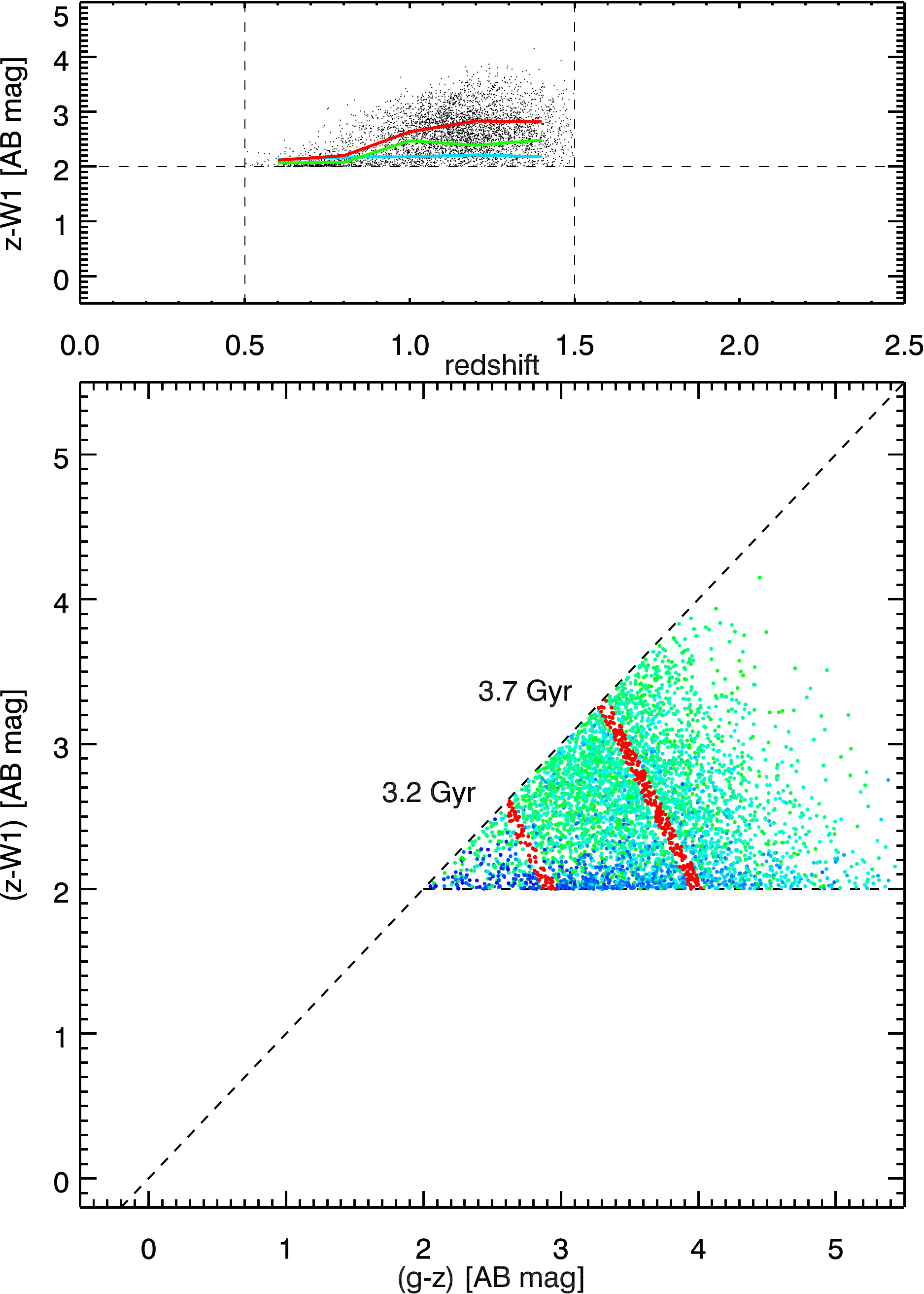}}
\caption{\small (Left) These two plots show expected galaxy tracks according to models from \citet{Bruzual2003}. The bottom plot shows the color-color selection of our $gzW1$ diagram in analogy to the $BzK$ selection of \citet[][]{Daddi2004}, with dashed lines corresponding to Equations (\ref{eq:gzw1}) and (\ref{eq:zw1}). Shown are tracks in redshift for fixed ages and assuming a star formation timescale $\tau \simeq 0.5$ Gyr. In the absence of extinction, our selection region will choose $\gtrsim 2$ Gyr population with $1 \lesssim z \lesssim 2$. The top plot shows the $z-W1$ evolution as a function of age and redshift. (Right) The same plots as on the left with the same scales and colors, but with our actual data. The colored lines in the top plot represent mean $z-W1$ values for 0.1-width redshift bins for each age. The red regions in the bottom plot represent roughly the slopes of the age lines in the left plot, and the ages given are mean ages for each red region.}
\label{fig:gzw1}
\end{figure*}

\begin{figure*}[ht]
\centerline{\includegraphics[height=4cm]{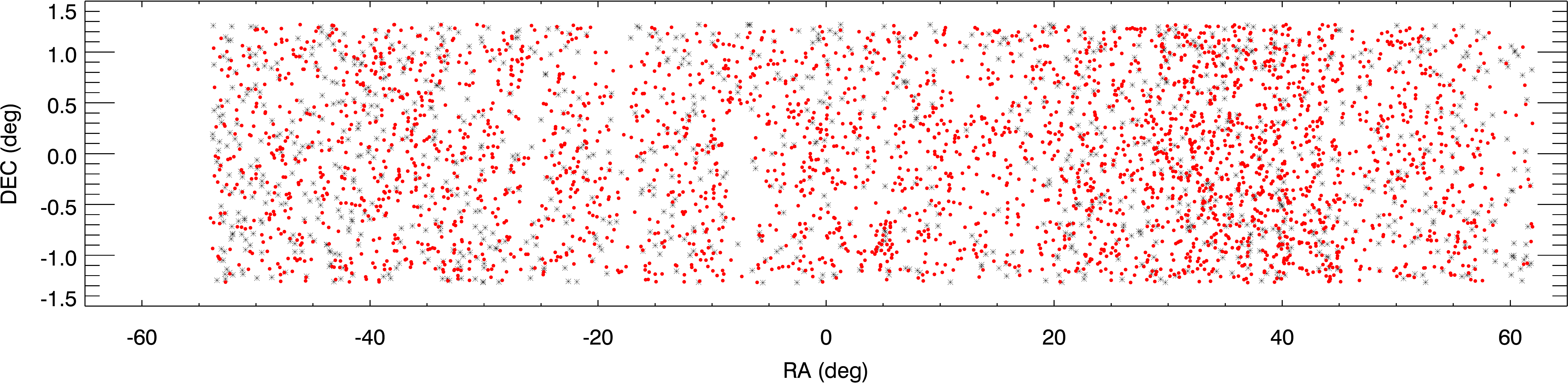}}
\caption{\small Location on the sky of our final selection of galaxies. Black represents $0.5 \leq z \leq 1.0$ (1179 galaxies) and red represents $1.0 \leq z \leq 1.5$ (3274 galaxies).   Note that this image has been stretched vertically for clarity, as the true aspect ratio of the field is $\approx 1/60.$\vspace{3mm}}
\label{fig:skygals}
\end{figure*}


\section{Galaxy Selection and Characterization}
\label{sec:galaxies}


In order to select galaxies best suited for our tSZ measurements, we have followed the selection criteria in \citet{Spacek2016}. We therefore have restricted our attention to massive elliptical galaxies with redshifts 0.5 $\leq z \leq$ 1.5. Galaxies are initially selected from the \textit{WISE} catalog to have equatorial coordinates that lie within SDSS Stripe-82 ($306^\circ < \alpha_{2000} < 62^\circ$ and $-1.27^\circ < \delta_{2000} < 1.27^\circ$). A cut was made requiring S/N $> 5$ in both $W1$ and $W2$.

We incorporated the SDSS bands in order to perform a color selection analogous to the passive $BzK$ selection of \citet{Daddi2004} using $g - z$ and $z - W1$ colors. This $gzW1$-selection is illustrated for several example spectral energy distributions (SEDs) from \citet{Bruzual2003} in the left plots of Figure \ref{fig:gzw1}. The selection lines (in the AB magnitude system) are
\begin{equation} 
(z - W1) \leq (g - z) - 0.02,
\label{eq:gzw1}
\end{equation}
and
\begin{equation} 
(z - W1) \geq 2.0,
\label{eq:zw1}
\end{equation}
which define the wedge in the upper right of the bottom plots in Figure \ref{fig:gzw1}. The color-selection was necessary to reduce the sample to a useable size. Galaxies whose colors lie in this wedge should be old and passively evolving galaxies at $1 \lesssim z \lesssim 2$. Following the \textit{WISE} and $gzW1$ criteria, and selecting only \textit{WISE} sources with unique SDSS-DR12 matches, we arrived at a sample of $\approx$ 30,000 galaxies which were further pared down using redshift and SED parameters. We emphasize that we are after a highly reliable sample and are willing to sacrifice completeness in the interest of purity.

Photometric redshifts were computed with extinction-corrected SDSS $ugriz$ and \textit{WISE} $W1$ and $W2$ photometry using EAZY \citep{Brammer2008}. The $W3$ and $W4$ bands were omitted since they are not comparably deep and are dominated by a warm/hot dust component that obscures the stellar component of the target galaxies that we are after. We then applied the constraint that $0.5 \leq z \leq 1.5$. The seven-band SEDs were fit with \citet{Bruzual2003} exponentially declining star formation rate (with timescale $\tau$) models. We only used those objects with reliable SED fits by selecting galaxies with reduced $\chi^2 \leq 5$. We applied further selections based on the results of the SED fits by taking only galaxies with ages $\geq 1$ Gyr and specific star formation rates $\leq 0.01$ Gyr$^{-1}$. This ensured that we were choosing older galaxies that were not actively forming stars, especially excluding dusty starbursts which made it through the color-cuts. This gave a sample of $\approx$10,000 galaxies, which were then pruned of known contaminants to the tSZ signal.

\begin{figure*}[ht]
\centerline{\includegraphics[height=4.5cm]{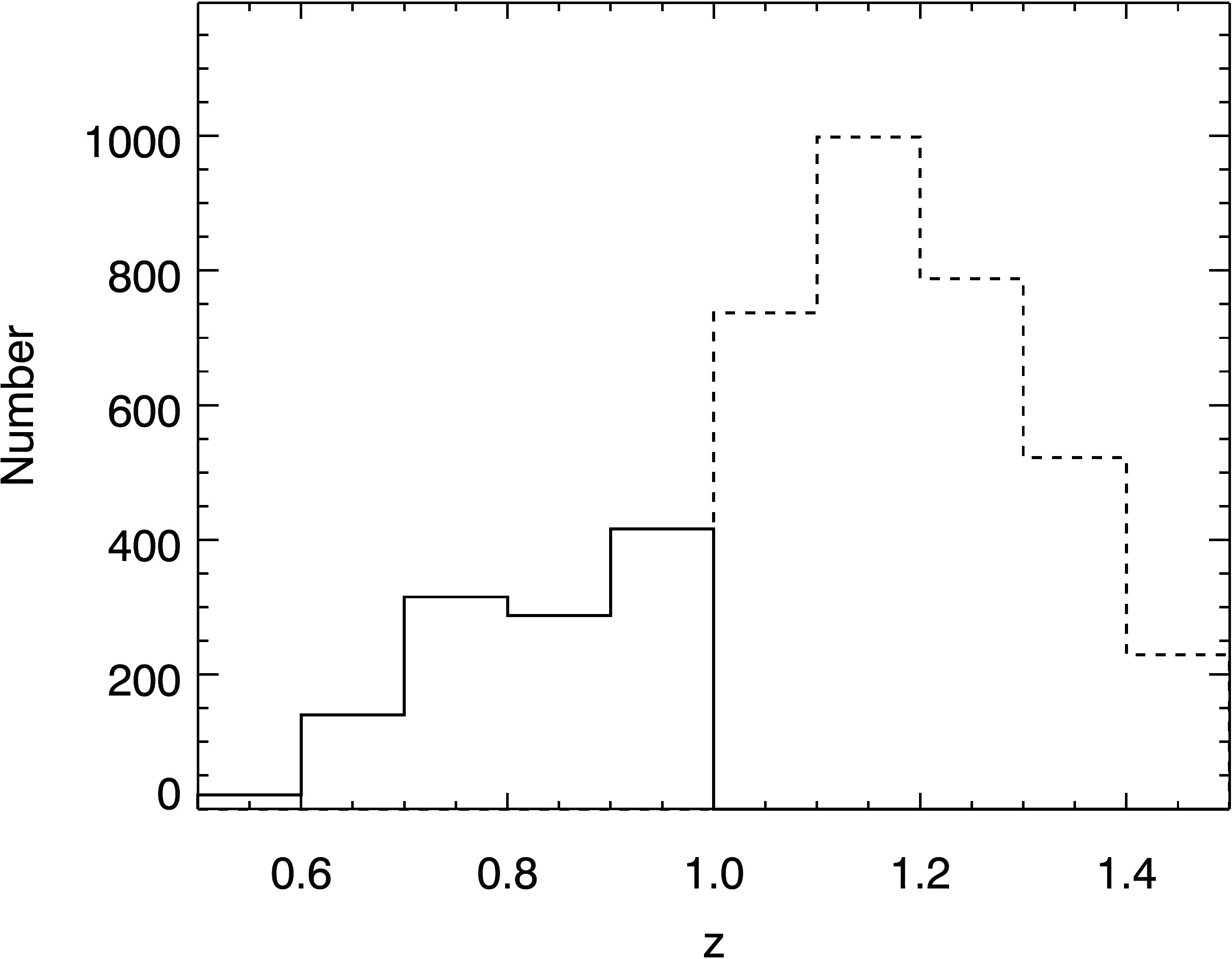}
\includegraphics[height=4.5cm]{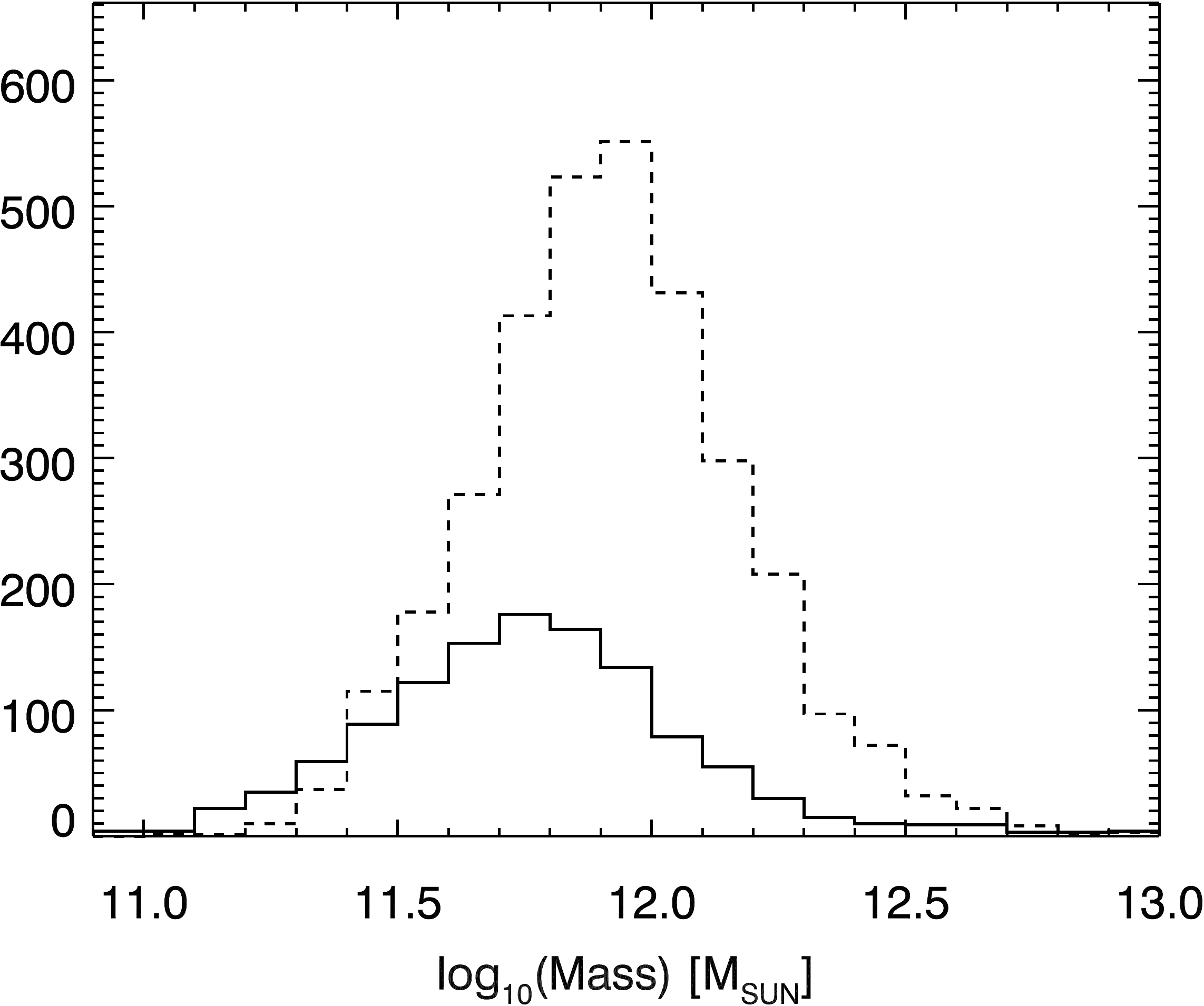}
\includegraphics[height=4.5cm]{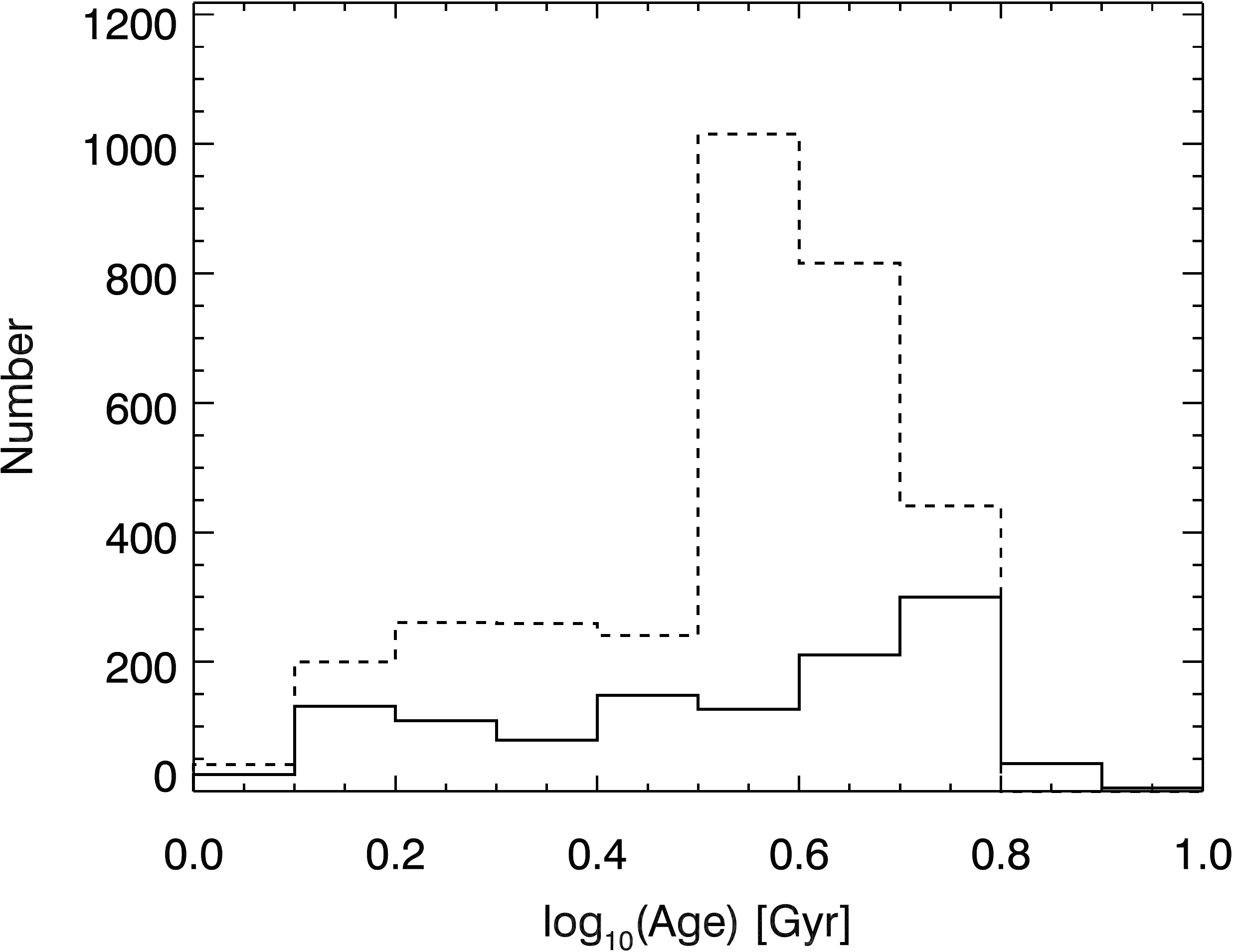}}
\caption{\small Redshift, mass, and age distributions of our final selection of galaxies. Solid lines represent $0.5 \leq z \leq 1.0$ (1179 galaxies) and dashed lines represent $1.0 \leq z \leq 1.5$ (3274 galaxies). The redshift histogram has a bin size of 0.1, and the mass and age histograms have bin sizes of 0.1 in log-space.}
\label{fig:zmass}
\end{figure*}

\begin{table*}[t]
\begin{center}
\resizebox{14cm}{!}{
\begin{tabular}{|c|c|c|c|c|c|c|c|c|c|}
	\hline
	Cut & $z$ & $N$ & $\left< z \right>$ & $\left< l_{\text{ang}}^2 \right>$ & $\left< \text{M} \right>$ & $\left< \text{Age} \right>$ & $\left< \text{L}_{W1} \right>$ & $\left< z \right> _\text{M}$ & $\left< l_{\text{ang}}^2 \right> _\text{M}$ \\
	&  & & & (Gpc$^2$) & (M$_\odot$) & (Gyr) & (erg s$^{-1}$ Hz$^{-1}$) &  & (Gpc$^2$) \\ \hline
	All & $0.5-1.0$ & 1179 & 0.83 & 2.56 & 7.81 $\times$ 10$^{11}$ & 3.80 & 7.83 $\times 10^{30}$ & 0.86 & 2.61 \\
	All &  $1.0-1.5$ & 3274 & 1.20 & 3.04 & 10.1 $\times$ 10$^{11}$ & 3.56 & 12.8 $\times 10^{30}$ & 1.21 & 3.05 \\
	Planck & $0.5-1.0$ & 227 & 0.83 & 2.55 & 6.93 $\times$ 10$^{11}$ & 3.63 & 7.04 $\times 10^{30}$ & 0.86 & 2.60 \\
	Planck &  $1.0-1.5$ & 529 & 1.21 & 3.05 & 9.68 $\times$ 10$^{11}$ & 3.44 & 12.4 $\times 10^{30}$ & 1.21 & 3.05 \\ \hline
\end{tabular}
}
\end{center}
\caption{\small Mean and mass-averaged values for several relevant galaxy parameters in the two final redshift ranges. ``All'' represents our complete, final galaxy sample, and ``Planck'' represents our final galaxy sample with further cuts applied, as discussed in Section \ref{sec:planck}. \vspace{2mm}}
\label{tab:meanvals}
\end{table*}

In order to estimate the possible contamination of our sample, we can appeal to morphological measurements where the same SED selection criteria as above are applied. Unfortunately, we cannot use the sample we have selected from Stripe-82 because Hubble Space Telescope (HST) resolution is required to classify galaxies at $0.5\lesssim z\lesssim1.5$. Instead we use the largest HST survey available, CANDELS \citep{Grogin2011,Koekemoer2011}. We use the SED fit parameters from \citet{Skelton2014} and Sersic function fits from \citet{vanderWel2012}, who used GALFIT \citep{Peng2002} to fit the 2D light profiles to the CANDELS HST images in the $H_{F160W}$-band. Since galaxies with stellar mass greater than $10^{11}$$M_{\sun}$ are rare on the sky and even the largest HST survey is significantly smaller than the $\approx$300 deg$^2$ Stripe-82 survey, we lower our mass limit to $M\geq10^{10}$$M_{\sun}$ for this exercise. We choose all galaxies with age greater than 1 Gyr and SSFR $< 0.01$ Gyr$^{-1}$. Additionally, we require a reliable Sersic fit from\\ \citet{vanderWel2012} and choose Sersic index $n > 2.5$ to be representative of ``elliptical'' galaxies. For samples of 364 and 346 galaxies, we find that 82\% and 78\% are $n > 2.5$ galaxies at $0.5\leq z \leq1.0$ and $1.0\leq z \leq 1.5$, respectively. If we remove the age and SSFR requirements we get only 45\% and 30\% $n > 2.5$ ellipticals at low and high redshift, respectively. Therefore, assuming morphology and SED  parameters are correlated in this way, we estimate that our sample of old galaxies that are not actively forming stars is $\simeq80$\% pure. Removing this 20\% contamination is part of the focus of Sections \ref{sec:contaminant} and \ref{sec:planck}.

The most significant contaminants that must be removed are known AGN and galaxy clusters. We therefore removed sources from the ROSAT Bright and Faint Source catalogs \citep[BSC and FSC;][]{Voges1999}. We additionally removed known clusters from ROSAT \citep{Piffaretti2011}. Clusters selected via the SZ effect would completely counter our measurements, so we removed both known \textit{Planck} \citep{PlanckCollaboration2015b} and ACT \citep{Marriage2011,Hasselfield2013} clusters.  X-ray sources from \textit{XMM-Newton} and \textit{Chandra} \citep{LaMassa2013}, sources from the AKARI/FIS Bright Source Catalog\\ \citep{Yamamura2010}, and sources from the AKARI/IRC Point Source Catalog \citep{Ishihara2010} were also removed. We also removed Galactic molecular clouds by cross-matching with the Planck Catalogue of Galactic Cold Clumps \citep{PlanckCollaboration2015c}, compact sources from the nine-band Planck Catalog of Compact Sources \citep{PlanckCollaboration2014}, and removed all sources from the IRAS Point Source Catalog \citep[][pp. 1-265]{IRAS1988} and radio sources from \cite{Best2012}. We also verified that none of our sources satisfied the ``$W1W2$-dropout'' criteria for extremely luminous infrared galaxies of \citet{Eisenhardt2012}. In all cases, sources with a possible contaminant within 4.0 arcmin, approximately double our region of interest around each source, were flagged and those sources were removed from further consideration. This left $\approx$ 7200 massive, quiescent, $0.5 \leq z \leq 1.5$ galaxies that are away from known potential contaminants.

Finally, to make sure we were selecting galaxies with the most reliable parameters, we limited the AB magnitude errors in the SDSS bands ($ugriz$), with $\text{mag\_error} < 1.5$ mag, required $\log_{10}(\text{SSFR})$ to be finite, and limited the galaxy stellar mass as $M < 10^{13} M_{\odot}$. This resulted in a final selection of 4453 galaxies to include in our tSZ stacks. To narrow down our measurements in redshift space we split our galaxy sample into two redshift bins: a ``low-$z$'' bin with 1179 $0.5 \leq z < 1.0$ galaxies and a ``high-$z$'' bin with 3274 $1.0 \leq z \leq 1.5$ galaxies. We show how our final galaxy selection fits in with our original SED color selection in the right plots of Figure \ref{fig:gzw1}. The locations of the final selection of galaxies is shown in Figure \ref{fig:skygals}. The mass, redshift, and age distribution of the final sample is shown in Figure \ref{fig:zmass}. Mean and mass-averaged values for redshift, angular diameter distance, mass, age, and $W1$ luminosity in both redshift bins are given in Table \ref{tab:meanvals}.


\section{Filtering}
\label{sec:filter}

Before stacking the ACT data around our selected galaxies, we needed to filter the ACT maps to remove the primary CMB anisotropy and maximize the signal-to-noise at the spatial scales we are measuring. An ideal Fourier-space point source filter is given by $\psi = \frac{\tau}{P} \left[ \int d^2k \frac{\tau^2}{P} \right]^{-1}$, where $\tau$ is the Fourier-space source profile and $P$ is the Fourier-space noise covariance matrix \citep[e.g.][]{Haehnelt1996}. For the source profile we assumed a slightly extended source such that $\tau = B \times G$, where $B$ is the Fourier-space beam function and $G$ is a Fourier-space Gaussian function. We have also approximated the noise $P$ as the ACT noise power spectra given in \citet{Das2014} for seasons 3 and 4 plus the CMB power spectrum. We therefore have approximated the filter as

\begin{equation}
\psi \approx \frac{B \times G}{N} \left[ \int d^2k \frac{(B \times G)^2}{N} \right]^{-1},
\label{eq:psi}
\end{equation}
where $N$ is the Fourier-space CMB+noise power spectrum. We did this for each band in each season and then averaged the two seasons together, resulting in an averaged filter for each band. For $G$ we chose a Gaussian with a FWHM of 1.5 arcmin to represent a slightly extended source. This is because our signal of interest is from hot gas within and surrounding the galaxies that likely represents the cumulative heating due to multiple cycles of AGN activity. We have no way of knowing the true shape of this gas but we expect the tSZ signal to be greatest near the galaxy and decrease away from it, meaning it is simplest to assume a slightly extended Gaussian profile. We note that this Gaussian we use is slightly larger than the ACT 148 GHz beam, which has a FWHM of 1.44 arcmin.

\begin{figure}[t]
\centerline{\includegraphics[height=6cm]{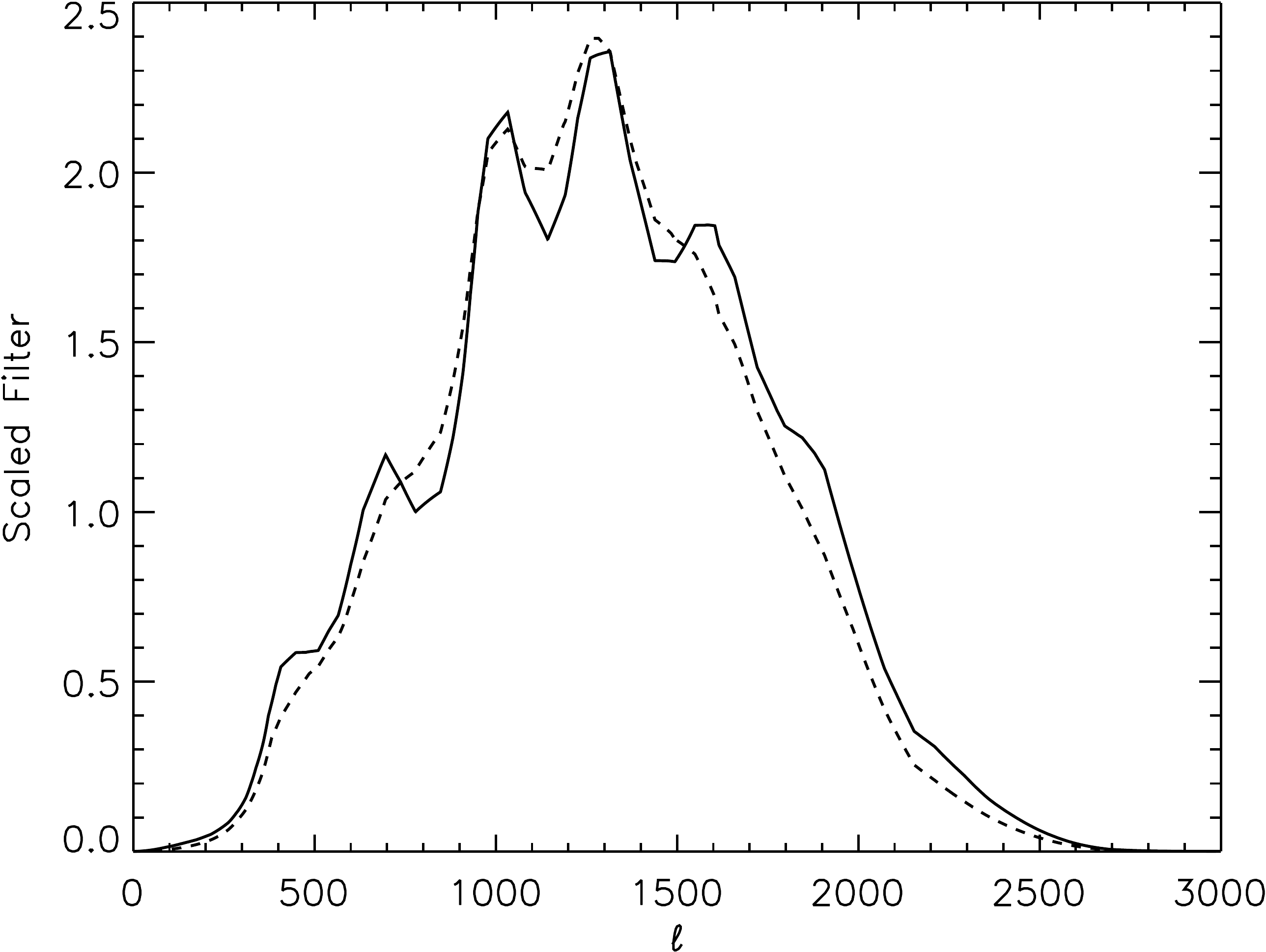}}
\caption{\small Scaled filters for both bands, averaged between seasons, in Fourier-space. The solid line represents 148 GHz and the dashed line represents 220 GHz. \vspace{3mm}}
\label{fig:filters}
\end{figure}

We then scaled the filters so that the flux within a 1 arcmin radius aperture is preserved in our maps after filtering, representing the regions of interest we measure around the selected galaxies. This choice of aperture follows from \citet{Spacek2016}, where it is noted that the energy input from AGN feedback is unlikely to affect scales much larger than twice the dark matter halo virial radius, which corresponds to about 2 arcmin at the redshifts we are investigating. Although the gas surrounding the galaxies will have both intrinsically different angular sizes and different angular sizes due to their differing redshifts, we have no way of knowing these sizes and the best we can do is take an aperture that is not unnecessarily large and that we expect to contain most of the signal in all cases. Additionally, we want these measurements to be comparable with cosmological AGN feedback simulations, and that is easiest to do with a constant measurement aperture. The factors used to scale the filters are 0.0167 and 0.0162 for 148 and 220 GHz, respectively. The final averaged, scaled filters are applied to the corresponding ACT maps for both seasons. Pictures of the final scaled Fourier-space filters are shown in Figure \ref{fig:filters}. The lack of smoothness is due primarily to the CMB power spectrum.

\section{Stacking}
\label{sec:stack}

\begin{figure*}[ht]
\centerline{\includegraphics[height=6cm]{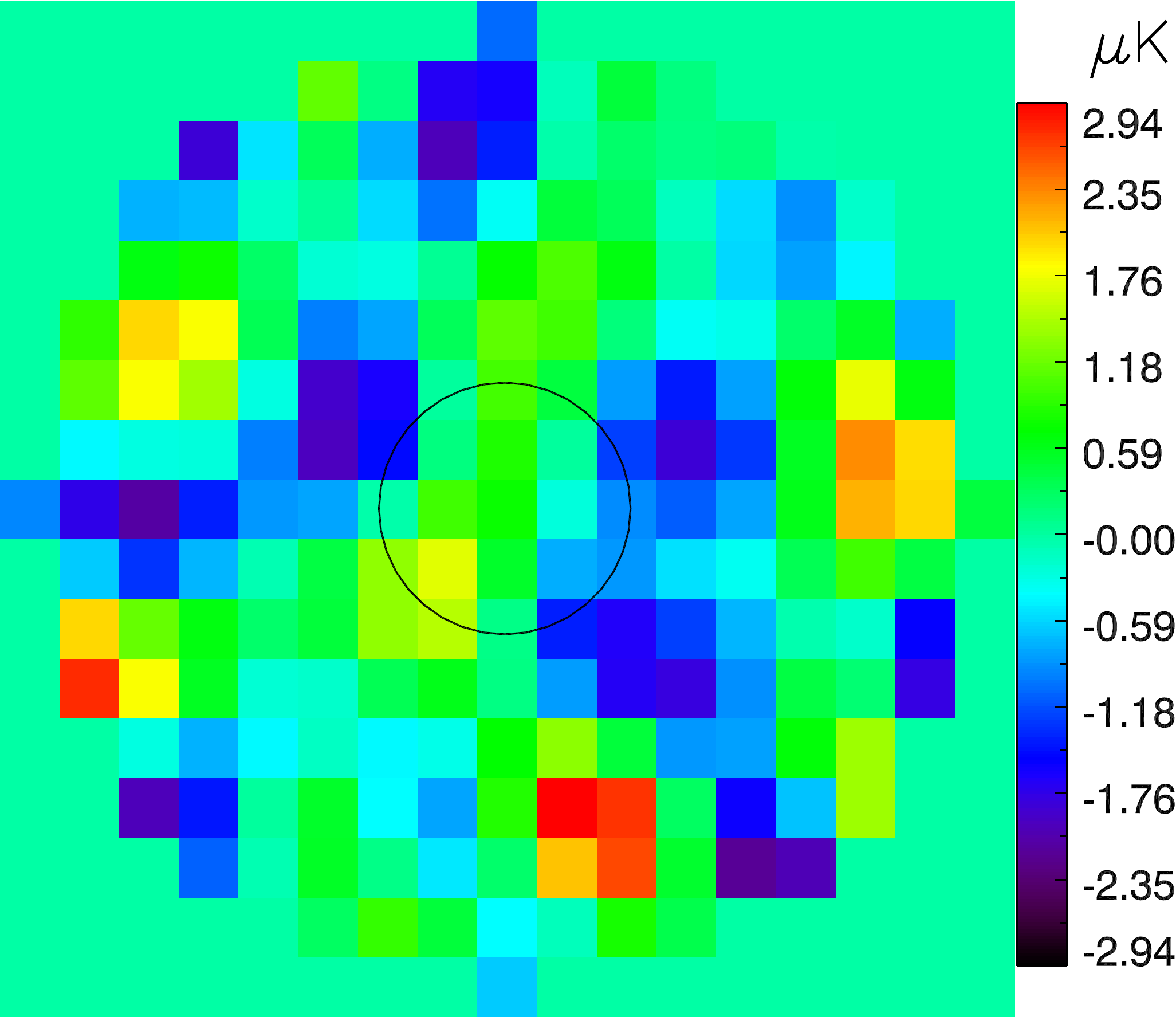}
\ \includegraphics[height=6cm]{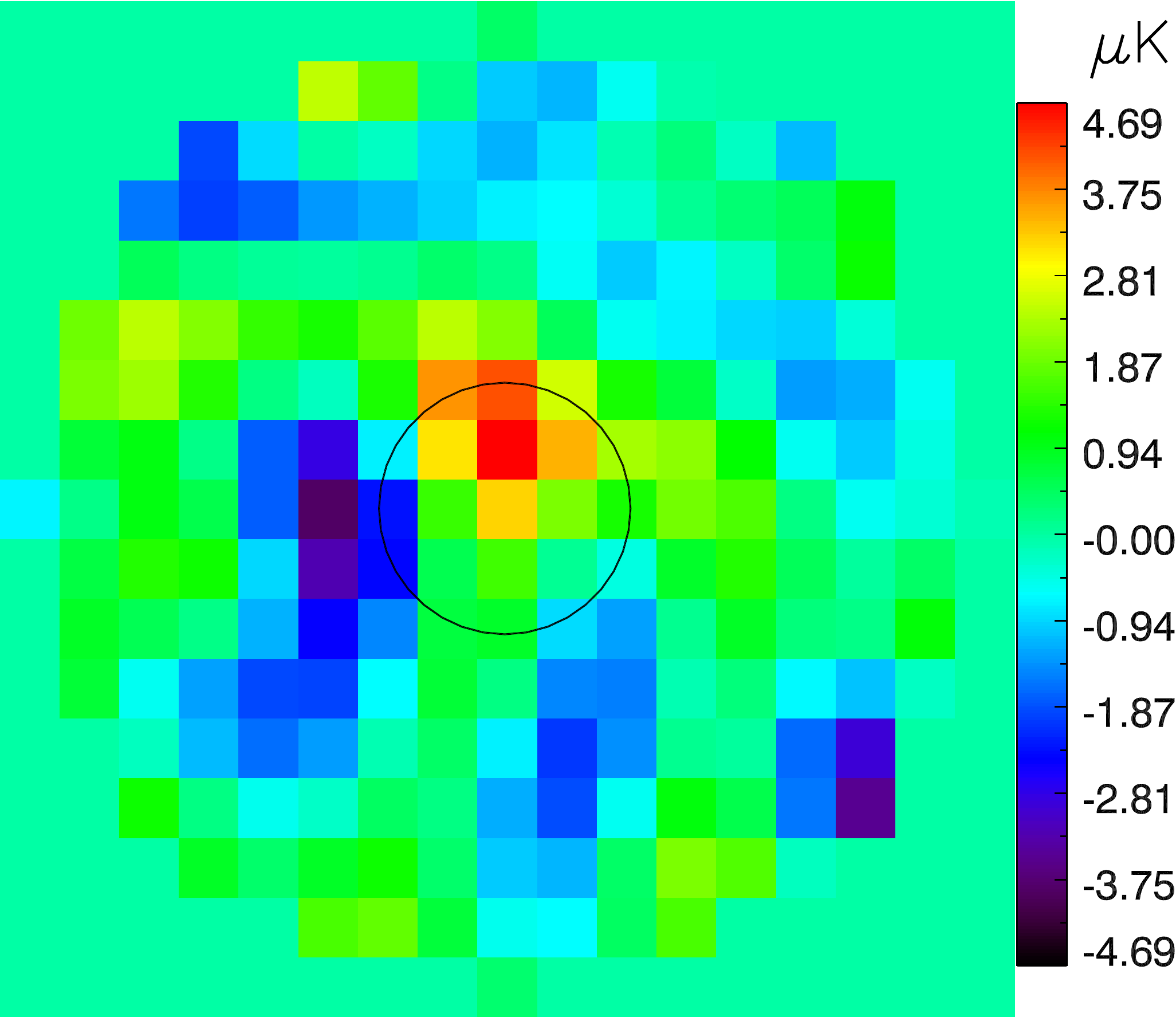}}
\vspace{2mm}
\centerline{\includegraphics[height=6cm]{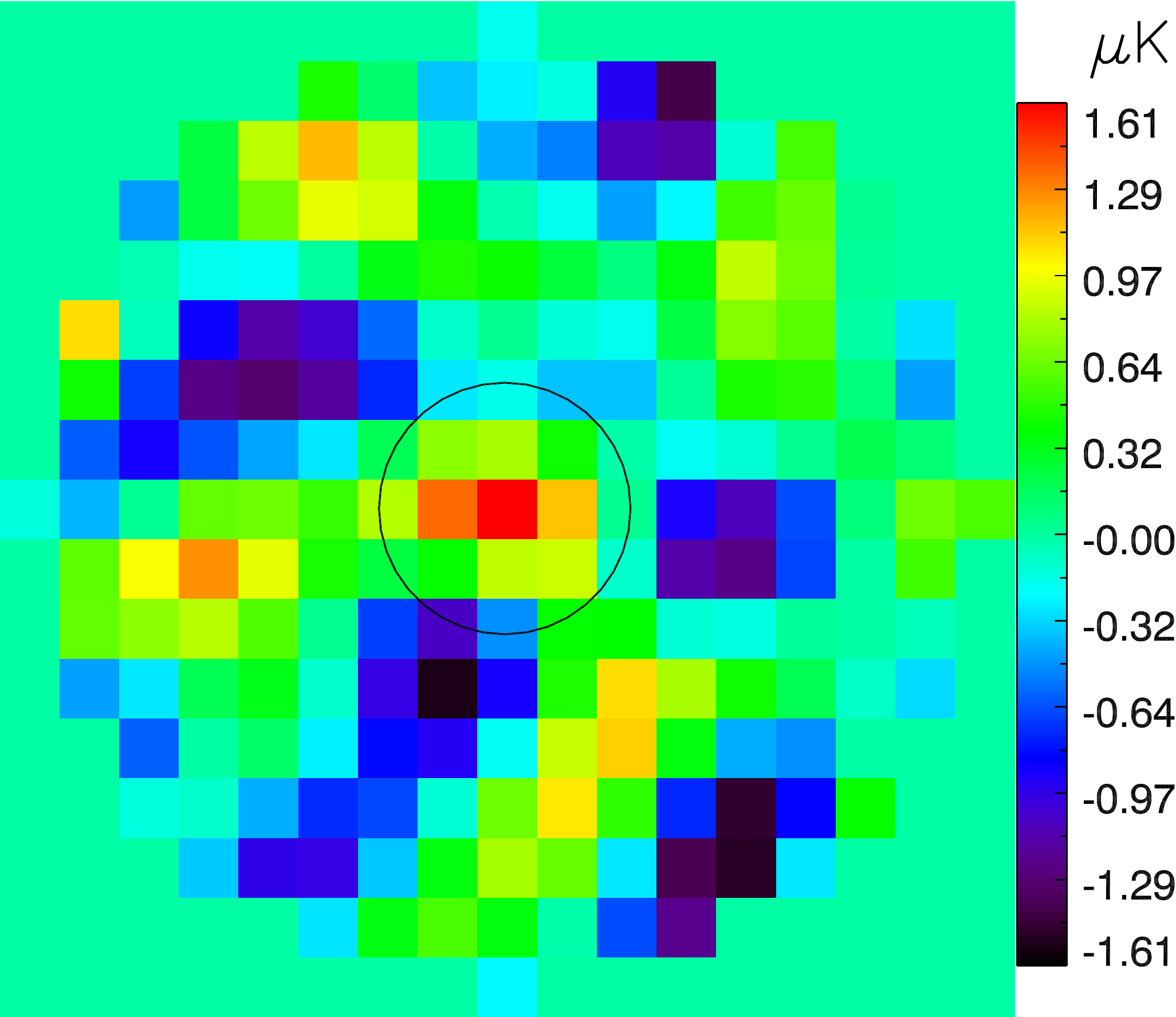}
\ \includegraphics[height=6cm]{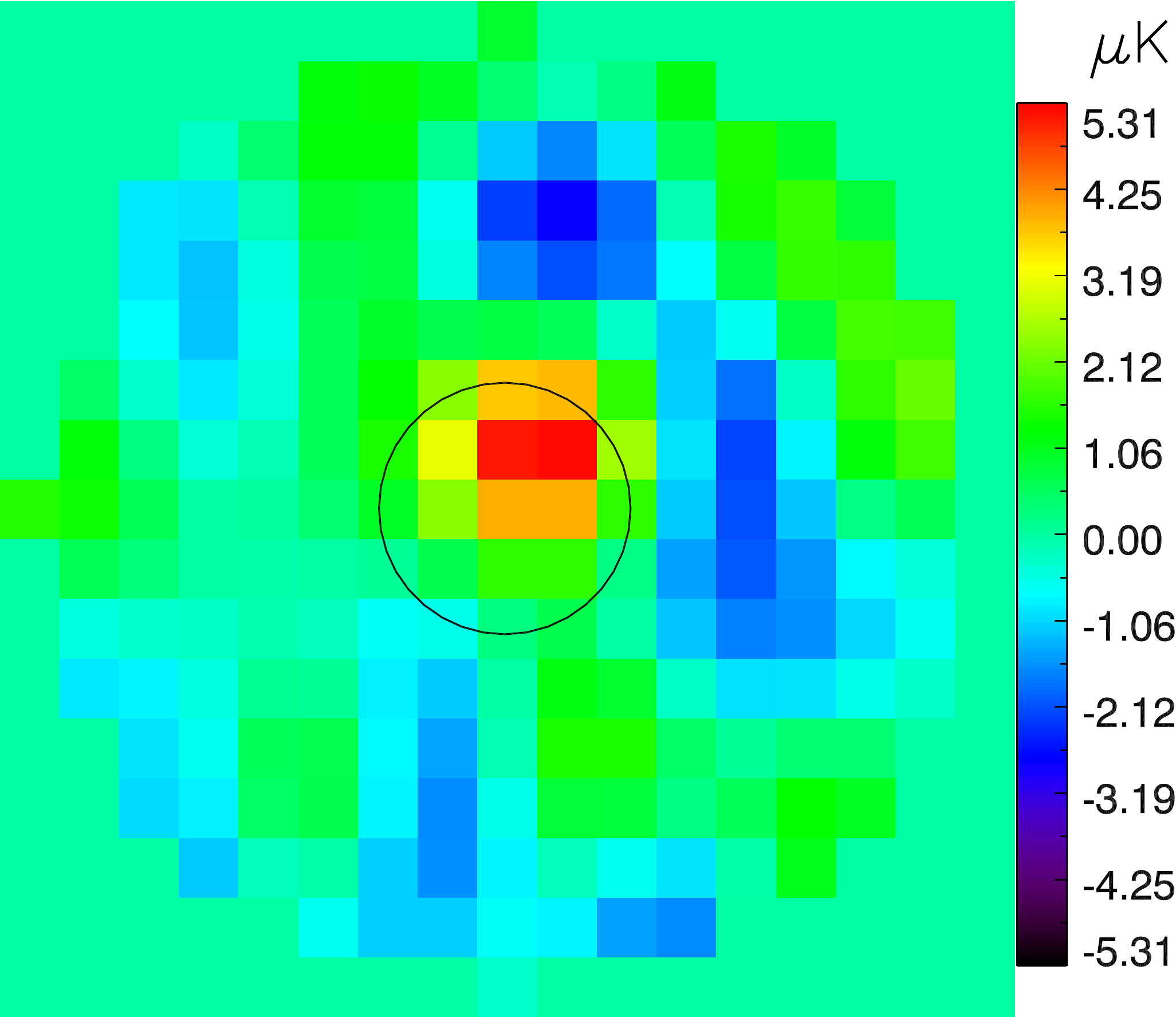}}
\caption{\small Season-averaged stacked galaxy stamps. Left is 148 GHz, right is 220 GHz, top is low-$z$ (1179 galaxies), bottom is high-$z$ (3274 galaxies). Units are $\mu$K, with black circles representing the 1 arcmin radius aperture we use for our final values. \vspace{3mm}}
\label{fig:firststamps}
\end{figure*}

To stack the CMB data we first made a 8.4$\times$8.4 arcmin (17$\times$17 pixel) stamp around each galaxy at 148 and 220 GHz in the filtered ACT data for seasons 3 and 4. Then we averaged the individual stamps together to make two stacked stamps for each band in each season, split into low-$z$ and high-$z$ galaxies. Finally, we averaged the seasons together. The resulting stamps are shown in Figure \ref{fig:firststamps}, with scales centered around 0 in units of $\mu$K. Any pixels $>$ 4 arcmin away from the center were set to 0 since that was the distance of our potential contaminant cuts. We get our final measurements integrated over the sky by summing the stacked signal within a 1 arcmin radius aperture (corresponding to a 2 pixel radius), shown as black circles in Figure \ref{fig:firststamps}.

The upper left panel of this figure shows a signal close to zero, while the lower left panel shows a clear positive signal in the center. These are the low-$z$ and high-$z$ 148 GHz stamps, respectively, and there is no tSZ detection in our initial stacks, which would be a negative signal at 148 GHz. In fact, at least at high-$z$, there is a significant contaminant signal. Looking at the right panels, at 220 GHz, we see that the stamps are even more dominated by positive contaminant signals. Since the tSZ effect has a negligible impact at this frequency, indicated by Equation (\ref{eq:newys}), this indicates that our galaxy selection process was imperfect, and there still remains a positive contaminating signal composed of faint sources that we were unable to account for. Looking at a typical range of emission by dust at $z=1$ (light and dark blue curves in Figure \ref{fig:allfilters}), the CMB spectrum (green curve in Figure \ref{fig:allfilters}), and the ACT bands (rightmost red hatched region in Figure \ref{fig:allfilters}), it seems likely that this contaminating signal at 220 GHz also extends into the 148 GHz band. It is therefore likely that we are in fact seeing a significant tSZ signal that is obscured by contaminant emission.

In order to estimate the uncertainty in our final measurements we generated 429,571 random points in our field on the sky, a number chosen by dividing the area of our field by the area of the 148 GHz beam which we approximated as $2\pi\sigma^2$, where $\sigma$ is the Gaussian beam standard deviation. We then applied the same 4 arcmin contaminant source cuts as we applied to our galaxy selection, leaving us with 294,176 random points. We stacked these random points on the sky for each band in the same way as we stacked our galaxies, and we computed the corresponding 1 arcmin radius aperture sums. First we computed an overall offset from the random points by getting the mean value of these sums. Since we remove galaxies anywhere near potential contaminants that might have positive or negative signal, we inherently bias the zeropoint of the ACT maps. We therefore corrected our aperture sums by subtracting off these mean offset values. These season-averaged offsets are 0.014 and -0.35 $\mu$K arcmin$^2$ for 148 and 220 GHz, respectively. Next we used the random point sums to compute the variance for an individual measurement in each case. If we first define the normal variance of $N$ random values $x$ as
\begin{equation}
\text{var} = \frac{1}{N-1} \sum_{i=1}^N\left( x_i - \left<x\right> \right)^2,
\label{eq:stddev}
\end{equation}
where $N=$ 294,176, we can then define our overall uncertainty $\sigma$ as a combination of measurement error and offset error, given by
\begin{equation}
\sigma = \left( \frac{\text{var}}{n} + \frac{\text{var}}{N} \right)^{1/2},
\label{eq:sigma}
\end{equation}
where $n$ is the number of galaxy measurements (1179 for low-$z$ and 3274 for high-$z$). The result is final co-added sums and uncertainties for each individual season-averaged band and redshift bin, and these are given in Table\ \ref{tab:coadds}.

From this table, we can directly see that there is a $\approx 1-2\sigma$ contaminant signal at 148 GHz, and a $\approx 3-6\sigma$ contaminant signal at 220 GHz. It is clear that obtaining the best possible constraints on non-gravitational heating and AGN feedback requires making the best possible separation between the tSZ signal and the contaminating signal, which is addressed in the following sections. Finally, we convert our co-added $\Delta T$ signal into gas thermal energy using Equation (\ref{eq:newEthrm}). These values are shown in Table\ \ref{tab:finvals}, under ``Data only.''


\section{Modeling and Removing Dusty Contamination}
\label{sec:contaminant}

\begin{figure*}[ht]
\centerline{\includegraphics[height=7cm]{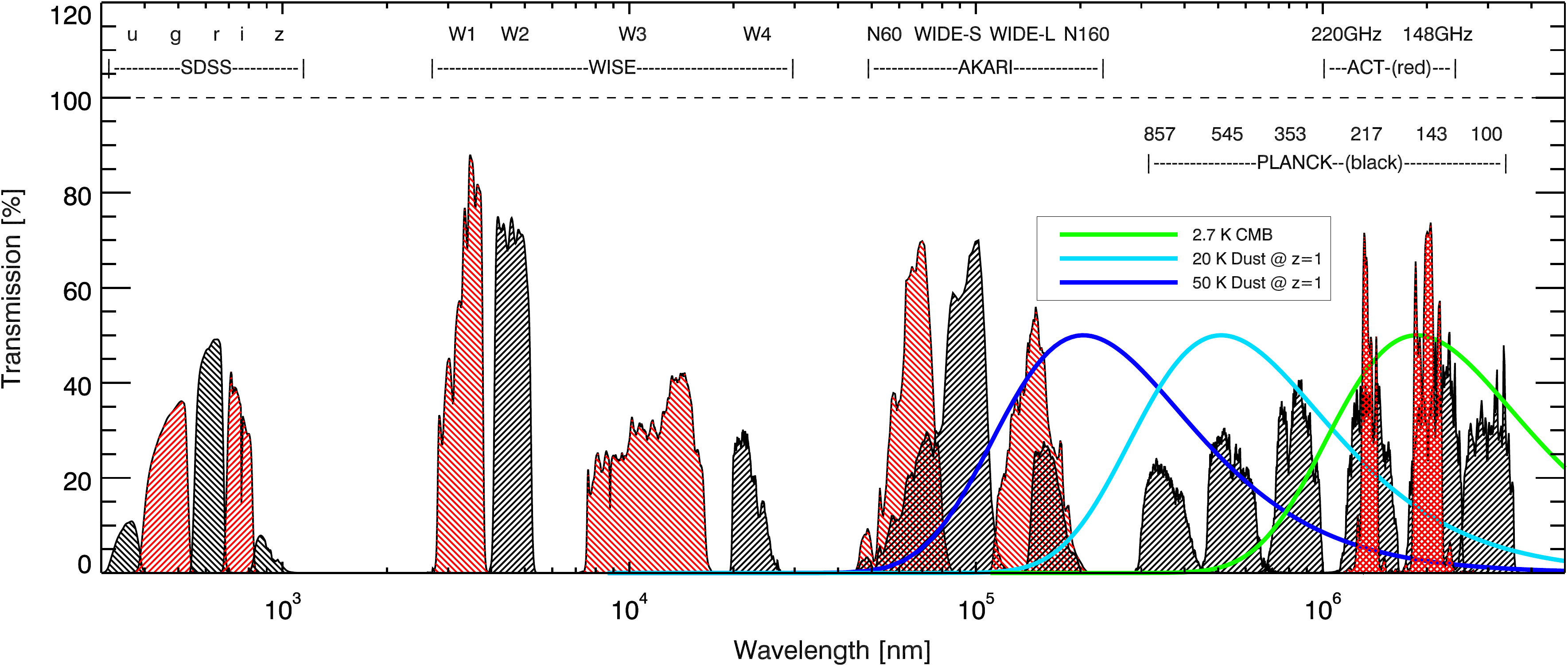}}
\caption{\small The filter curves for several of the data sets used in this paper. From left to right: SDSS and \textit{WISE} bands used for galaxy selection, AKARI and \textit{Planck} bands used for identifying and constraining the signal from dusty contaminating sources, and ACT bands used for measuring the tSZ effect. The first three surveys alternate between black and red for each band for clarity, while Planck bands are all black and ACT bands are all red to distinguish between the two. Also shown are blackbody curves for the CMB (green), 20 K dust at $z = 1$ (light blue), and 50 K dust at $z = 1$ (dark blue), all normalized to 50\% on the plot. The horizontal dashed line indicates 100\% transmission. \vspace{5mm}}
\label{fig:allfilters}
\end{figure*}

As evidenced by Table \ref{tab:coadds}, there appears to still be a significant contamination signal indicated by the large positive 220 GHz values, which is likely contributing to the 148 GHz values that we are interested in for our tSZ measurements. This is illustrated by the blue lines in Figure \ref{fig:allfilters}, where it is clear that dust at reasonable temperatures around $z=1$ will have significant emission in the ACT bands we are using. In order to constrain and subtract out this undetected contamination, we have followed the process described in \citet{Spacek2016} and built a detailed model of contaminants based on extrapolations of the source counts measured for SPT data in \citet{Mocanu2013}. We extended these source counts to fainter values by modeling a random population of undetected sources that follow the trend of the detected sources into the unresolved region. We then related these models to the contaminating signal in our 148 and 220 GHz measurements.

Following \citet{Mocanu2013} we separated contaminants into synchrotron sources, which emit mostly at lower frequencies, and 
dusty sources, which emit mostly at higher frequencies. For each source population we modeled the number counts as a power law, 
\begin{equation}
\frac{dN}{dS} = \frac{N_0}{S_{\rm max}} \left(\frac{S}{S_{\rm max}}\right)^{\alpha},
\label{eq:dnds}
\end{equation}
where $dN/dS$ is the number of sources between flux $S$ and $S + dS$, $N_0$ is an overall amplitude, $\alpha$ is the power-law slope,
and $S_{\rm max}$ is the flux at which we expect all brighter sources to have a 100\% completeness level in the source count catalog.
 We then computed a range of allowed source count slopes from the \citet{Mocanu2013} data,  by carrying out a $\chi^2$ fit in log-space.
 Our best-fit slopes at 220 GHz were $\alpha_s = -2.08 \pm 0.09$ for the synchrotron sources and $\alpha_d = -2.91 \pm 0.17$ for the dusty sources.

Note that our calculated values for $\alpha_d$ are much steeper than $\alpha_s$, meaning that while the number density of detected sources is dominated by synchrotron sources, the number density of undetected sources is likely to be dominated by dusty emitters.   Note also that $\alpha_d$  and $\alpha_s$  are sufficiently steep that the number of sources diverges as $S$ goes to 0, meaning that the source count distribution must fall off below some as-yet undetected flux.   For simplicity, we modeled this fall-off as a minimum flux $S_{\rm min}$ below which there are no contaminating sources associated with the galaxies we are stacking.  

\begin{table}[t]
\begin{center}
\resizebox{7cm}{!}{
\begin{tabular}{|c|c|c|}
\hline
Redshift & Band & 1-arcmin-radius sum\\
$z$ & (GHz)  &  ($\mu$K arcmin$^2$) \\ \hline
0.5 - 1.0 & 148  & 1.0 $\pm$ 1.4 \\              
0.5 - 1.0 & 220  & 6.2  $\pm$ 2.3 \\     
1.0 - 1.5  & 148 & 2.1  $\pm$ 0.9 \\    
1.0 - 1.5 & 220  & 8.7   $\pm$ 1.4 \\ \hline
\end{tabular} }
\end{center}
\caption{\small Final season-averaged co-added signals. The columns show redshift bin, band, and integration over a 1 arcmin radius region around the galaxies. \vspace{1mm}}
\label{tab:coadds}
\end{table}

For any choice of $\alpha_d,$ $\alpha_s,$ and $S_{\rm min}$ (which we will call a ``source-count model''), we are then able to construct a model population of contaminating source fluxes through a four-step procedure as follows.
(i) For each model source, we randomly decided whether it is a synchrotron source or a dusty source, such that the overall fraction of detectable dusty sources to synchrotron sources matches the observed source counts. We used a maximum flux cutoff of 305.7 $\mu$K arcmin$^2$, corresponding to the faintest bin of detected sources found by \citet{Mocanu2013}.
(ii) We then assigned the source a random 220 GHz flux, $S_{\rm 220, rand},$ by inverting 
\begin{equation}
\int^{S_{\rm 220, rand}}_{S_{\rm220,  min}} dS \, \frac{dN}{dS} =  R  \int^{S_{\rm 220, max}}_{S_{\rm 220, min}} dS \, \frac{dN}{dS}, 
\end{equation}
where $R \in \left[0,1\right]$ is a random number, 
such that their overall population matched the source count slopes. This gives
\begin{equation}
S_{\rm 220, rand} = \left[ \left( 1-R \right) S_{\rm 220, min}^{\alpha+1} + R \, S_{\rm 220, max}^{\alpha+1}\right]^{\frac{1}{\alpha+1}}.
\label{eq:srand}
\end{equation}
(iii) To obtain a corresponding flux for the source at 148 GHz we used the $\alpha^{150}_{220}$ spectral index distributions from \citet{Mocanu2013}, which we assume to have normalized Gaussian shapes with the properties (center, $\sigma$) = (-0.55, 0.55) for synchrotron sources and (3.2, 0.89) for dusty sources. We then randomly chose $\alpha^{150}_{220}$ values that fit these distributions and calculated the 148 GHz flux \citep[following][]{Mocanu2013} as
\begin{equation}
S_{148, \text{rand}}=\frac{S_{220, \text{rand}}}{C_1 \times C_2},
\label{eq:s220}
\end{equation}
where $C_1$ is the conversion factor between Jy and $\mu$K arcmin$^2$ integrated over the band filter curves, $C_2$ is the relating factor $(\nu_{220}/\nu_{148})^{\alpha^{150}_{220}}$ integrated over the band filter curves, and we used units of $\mu$K arcmin$^2$ for all $S$. (iv) Finally, we estimated the completeness of our 220 GHz measurements and randomly discarded modeled sources to match the estimated fraction of 220 GHz sources detected per flux. To do the estimation, we assumed a cutoff $S_{\text{cut}}$ of 250 $\mu$K arcmin$^2$, representing the $3\sigma$ limit of our 220 GHz source measurements. We discarded galaxies with signals greater than $S_{\text{cut}}$ or signals less than $-S_{\text{cut}}$. We then determined the completeness fractions for the modeled sources, and accounted for our measurement uncertainty, by adding the 220 GHz random point distribution onto $S_{\text{cut}}$ in a cumulative manner using the fraction of random point measurements below a given flux. This means $S_{\text{cut}}$ has a 50\% completeness, while fainter fluxes are increasingly less complete and brighter fluxes are increasingly more complete.
 
 \begin{table*}[t]
\begin{center}
\resizebox{15cm}{!}{
\begin{tabular}{|c|c|c|c|c|c|c|c|}
\hline
Model & $N$ & $z$ & $\int \Delta T_{150}(\bm{\theta}) d\bm{\theta}$ & $Y$ & $E_\text{therm} (\pm1\sigma)$ & $E_\text{therm} (\pm2\sigma)$ & S/N \\ 
          &    &        &  ($\mu$K arcmin$^2$)                                         & ($10^{-7}$ Mpc$^2$) & ($10^{60}$ erg) & ($10^{60}$ erg) & ($E_\text{therm}/1\sigma$)  \\  \hline
Data only & 1179  & $0.5-1.0$ & $1.0 \pm 1.4$  & $-0.8 \pm 1.1$ & $-2.8 \pm 3.9$ & $-2.8 \pm 7.9$ & -0.72  \\              
 & 3274  & $1.0-1.5$ & $2.1 \pm 0.9$  & $-1.9 \pm 0.8$ & $-7.0 \pm 3.0$ & $-7.0\pm 6.0$ & -2.33  \\
$\chi^2$ (ACT only) & 1179 & $0.5-1.0$ & $-1.6_{-1.9}^{+1.8}$  & $1.2_{-1.4}^{+1.4}$ & $4.5_{-5.1}^{+5.4}$ & $4.5_{-10.7}^{+11.0}$ & 0.85  \\
 & 3274 & $1.0-1.5$ & $-2.1_{-1.2}^{+1.1}$  & $1.9_{-1.0}^{+1.1}$ & $7.0_{-3.7}^{+4.0}$ & $7.0_{-7.7}^{+8.4}$ & 1.78  \\
$\chi^2$ (With \textit{Planck}) & 227 & $0.5-1.0$ & $-2.0_{-2.1}^{+2.0}$  & $1.5_ {-1.5}^{+1.6}  $ & $5.6_{-5.6}^{+5.9}$ & $5.6_{-11.6}^{+12.4}$ & 0.97  \\
 & 529 & $1.0-1.5$ & $-2.1_{-1.4}^{+1.3}$  & $1.9_ {-1.2}^{+1.3}$ & $7.0_{-4.4}^{+4.7}$ & $7.0_{-9.4}^{+9.7}$ & 1.50  \\
\hline
\end{tabular} }
\end{center}
\caption{\small Our final tSZ measurements using various methods for removing contamination. The last three columns represent the best fit $E_\text{therm}$ values with $\pm1\sigma$ values and $\pm2\sigma$ values and the $E_\text{therm}$ signal-to-noise ratio ($E_\text{therm}/1\sigma$), respectively.}
\label{tab:finvals}
\end{table*}

For each source-count model, we repeated the process 100,000 times, resulting in a large catalog of contaminating fluxes in both bands. From these, we computed the mean flux per contaminating source in each band, $\left<S_{150,\text{cont}}\right>$ and $\left<S_{220,\text{cont}}\right>$, which represents the contamination we are measuring in our stacks. To account for variations in the input parameters, we computed model contamination signals for a wide range of source-count models, with $S_{220,\text{max}} = 305.7 \, \mu$K arcmin$^2$. We varied $\alpha_s$ from $-2.26$ to $-1.90$ in steps of $0.09$ and we varied $\alpha_d$ from $-3.25$ to $-2.57$ in steps of $0.17$, representing ranges of $\pm2\sigma$ in steps of $\sigma$. We let $\log_{10}$($S_{\rm min}$) vary from $\log_{10}$(0.01 $\mu$K arcmin$^2$) to $\log_{10}$(30 $\mu$K arcmin$^2$) in steps of 0.2 in log-space.

\begin{figure*}[ht]
\centerline{\includegraphics[height=10cm]{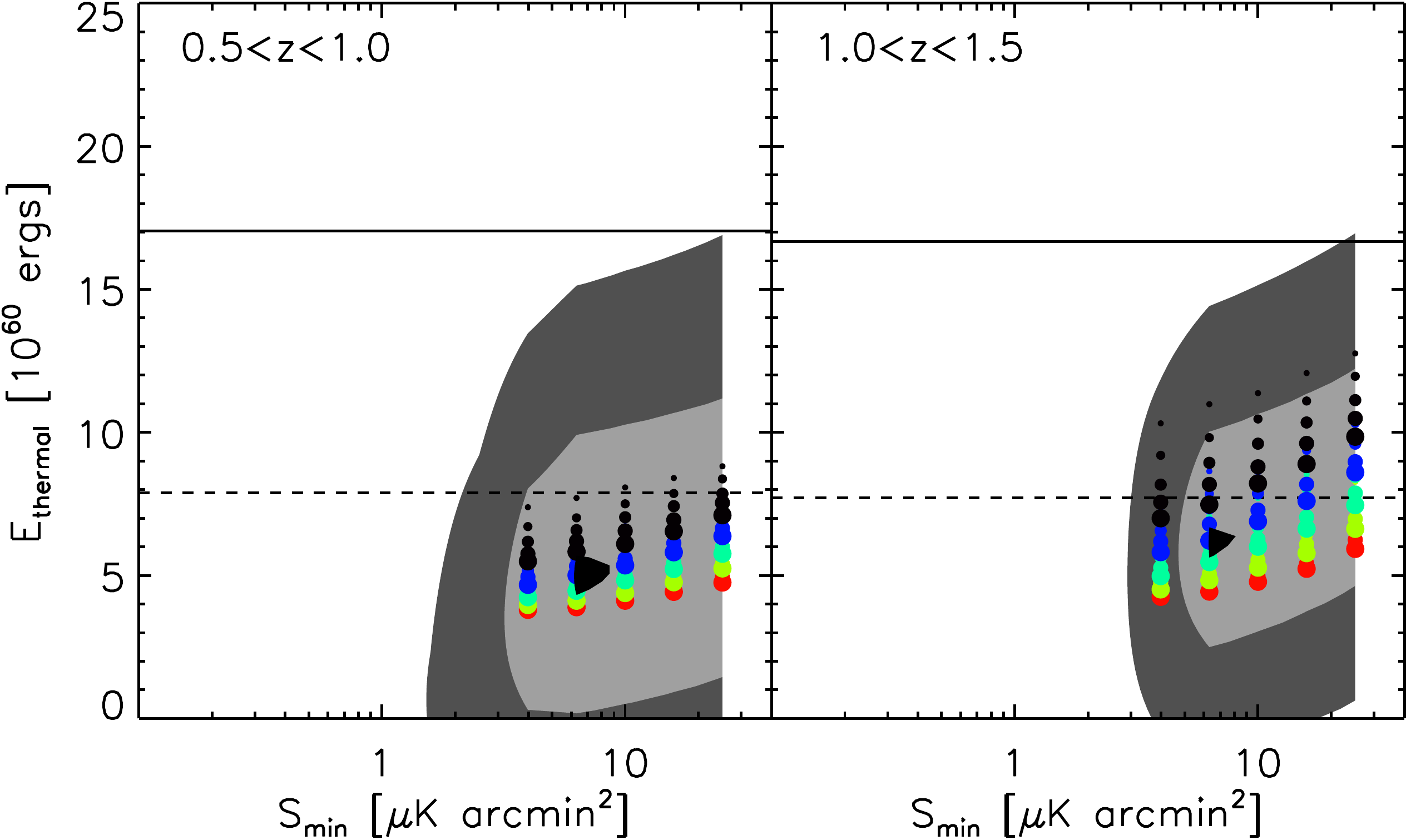}}
\caption{\small Plot of the contaminant-corrected $E_\text{therm}$ (see Equation (\ref{eq:newEthrm})) for different choices of $\alpha_{\rm dust}$, $\alpha_{\rm sync}$, and $S_{\rm min}$. Points are located at the peak $\chi^2$ probability for each model. Increasing size represents increasing (i.e. more positive) $\alpha_s$, and changing color from red to black represents increasing $\alpha_d$. The light and dark gray regions represent the complete span of $\pm1\sigma$ and $\pm2\sigma$, respectively, for all points. Black regions represent the most favorable models with peak $\chi^2$ probability. The horizontal solid black lines represent the best estimates for $E_\text{grav}$, and the horizontal dashed black lines represent the $-1\sigma$ values for $E_\text{grav}$ (see Equation (\ref{eq:Egrav})). \vspace{3mm}}
\label{fig:s148corr}
\end{figure*}

For each source-count model, we computed best-fit tSZ values by varying our two free parameters, tSZ signal ($S_{\text{SZ}}$) and the fraction of our measured sources that are contaminated ($f_{\text{cont}}$).   
We varied the tSZ signal from -50 to 50 $\mu$K arcmin$^2$ in steps of 0.1 $\mu$K arcmin$^2$, and we varied the fraction contaminated from -3 to 9 in steps of 0.01. For every combination of these parameters we computed a $\chi^2$ value,
\begin{equation}
\chi^2(f_{\text{cont}},S_{\text{SZ}}) = \mathcal{B} \times \mathcal{A}^{-1} \times \mathcal{B}^T ,
\label{eq:chi2matrixes}
\end{equation}
where $\mathcal{B}$ is the signal array,
\begin{equation}
\mathcal{B} = \left( \begin{array}{c}
f_{\text{cont}} \times \left<S_{150,\text{cont}}\right> + S_{\text{SZ}} - S_{150} \\
f_{\text{cont}} \times \left<S_{220,\text{cont}}\right> - S_{220}
\end{array} \right) ,
\label{eq:sigarray}
\end{equation}
and $\mathcal{A}$ is the noise matrix containing the noise for each band plus the covariance terms between each band,
\begin{equation}
\mathcal{A} = \left( \begin{array}{cc}
\sigma_{150}^2 & \sigma_{150}\sigma_{220} \\
\sigma_{150}\sigma_{220} & \sigma_{220}^2
\end{array} \right) .
\label{eq:noisetrix}
\end{equation}
Here, $S_{150}$, $S_{220}$, $\sigma_{150}$, and $\sigma_{220}$ are our measured 1 arcmin radius values from Table \ref{tab:coadds}. The $\sigma$ values are computed using random point measurements, given by
\begin{equation}
\begin{split}
& \sigma_i \sigma_j = \\ & \frac{\sum_{a=0}^{N_{\text{rand}}} (S^a_{i,\text{rand}}-\left<S_{i,\text{rand}}\right>)\times(S^a_{j,\text{rand}}-\left<S_{j,\text{rand}}\right>)}{N_{\text{rand}} N_{\text{source}}} ,
\label{eq:sigmaact}
\end{split}
\end{equation}
where $i$ and $j$ represent the bands, $S_{a,\text{rand}}$ and $S_{b,\text{rand}}$ represent the 1 arcmin radius aperture values for the random points, $N_{\text{rand}} = $ 294,176 is the number of random points used, and $N_{\text{source}}$ is the number of galaxies used (1179 for low-$z$ and 3274 for high-$z$). We then converted the $\chi^2$ values to Gaussian probabilities $P$ by taking
\begin{equation}
\begin{split}
& P(S_{\text{SZ}}) = \\ & \frac{\sum_{f_{\text{cont}} \in [0,1]}  \exp[-\chi^2(f_{\text{cont}},S_{\text{SZ}})/2]}
		                     {\sum_{f_{\text{cont}}} \sum_{S_{\text{SZ}}}  \exp[-\chi^2(f_{\text{cont}},S_{\text{SZ}})/2]}.
\label{eq:probmod}
\end{split}
\end{equation}
where the lower sum over $f_{\text{cont}}$ runs from $-3$ to $9$ and the lower sum over $S_{\text{SZ}}$ runs from $-50$ to $50$ $\mu$K arcmin$^2$. Our approach was thus to marginalize over values of $f_{\rm cont}$ in the full physical range from $0$ to $1,$ but normalize the overall probability by the sum of $f_{\text{cont}}$ over a much larger range, including unphysical values.  This excludes models in which a good fit to the data can only be achieved by moving $f_{\text{cont}}$ outside the range of physically possible values.

Equation (\ref{eq:probmod}) then gives us a function $P(S_{\text{SZ}})$ for each combination of $\alpha_d,$ $\alpha_s,$ and $S_{\rm min}$. We can convert the corresponding $S_{\text{SZ}}$ value to the gas thermal energy, $E_\text{therm},$ using Equation (\ref{eq:newEthrm}) and the average $l^2_\text{ang}$ from Table \ref{tab:meanvals}. Note that a positive detection of the tSZ effect is seen as a negative $\Delta T$ signal at 148 GHz, and it represents a positive injection of thermal energy into the gas around the galaxy. Additionally, we compute a corresponding range for $E_\text{grav}$ using Equation (\ref{eq:Egrav}) and values from Table \ref{tab:meanvals}. The peak of each $P(S_{\text{SZ}})$ curve is shown as the colored points in Figure \ref{fig:s148corr}, where $\alpha_s$ (represented by point size) is increasing (i.e.\ becoming more positive) downwards, and $\alpha_d$ (represented by point color) is increasing  upwards. The 1$\sigma$ and 2$\sigma$ contours are computed for each $S_{\rm min}$ by averaging $P(S_{\text{SZ}})$ across $\alpha_d$ and $\alpha_s.$  The resulting probability distribution depends only on $E_\text{therm}$ and $S_{\rm min}$, and 1$\sigma$ and 2$\sigma$ are represented by the values $P(S_{\text{SZ}})=0.61$ and 0.13, respectively (i.e. $\exp[-\sigma^2/2]$). These contours are shown in Figure \ref{fig:s148corr}, along with the $-1\sigma$ range for $E_\text{grav}$. Peak probability values are chosen using $P(S_{\text{SZ}})>0.99$, shown as the black regions in Figure \ref{fig:s148corr}. From this figure we see that there is a $\approx 1\sigma$ tSZ detection for $S_{\text{min}} \gtrsim 3 \mu$K arcmin$^2$ at low-$z$. At high-$z$ we see a $\approx 2\sigma$ tSZ detection for $S_{\text{min}} \gtrsim 5 \mu$K arcmin$^2$.

Finally, we average the probability distribution across $S_{\rm min}$ to get a final distribution as a function of only $E_\text{therm}$. The significance values of this curve are shown in Table\ \ref{tab:finvals} under ``ACT only.'' We see a $0.9\sigma$ tSZ detection at low-$z$ and a $1.8\sigma$ detection at high-$z$.


\vspace{5mm}
\section{Modeling and Removing Dusty Contamination With \textit{Planck}}
\label{sec:planck}

As indicated by Figure \ref{fig:allfilters}, we should be able to better constrain the contamination due to undetected dusty sources by incorporating data at higher frequencies than the ACT bands we are using to make our measurements. We therefore made use of the 2015 public data release from the \textit{Planck} mission, and focused on the high-frequency bands at 217, 353, 545, and 857 GHz. These data, with a $\approx$5 arcmin FWHM beam \citep{PlanckCollaboration2015d}, is too low-resolution to be useful in our direct tSZ measurements, but can still provide useful information about our galaxies at higher frequencies where the contaminant emission should be much brighter (see the light blue and dark blue curves in Figure \ref{fig:allfilters}). By incorporating these \textit{Planck} measurements we should be able to better discriminate between contaminant models, allowing us to better identify the true tSZ signal.

To utilize the \textit{Planck} data, we followed the same process as in the previous section by computing $\chi^2$ values for a number of modeled contaminants, but now we had several extra terms in each computed $\chi^2$ relating to the \textit{Planck} measurements. In order to stack our galaxies in the \textit{Planck} data, we first extended our contaminant source cut distance from 4 to 10 arcmin due to the much lower resolution. This resulted in a significant decrease in our number of galaxies, with 227 at low-$z$ and 529 at high-$z$. In order to filter out the primary CMB signal, we convolved each \textit{Planck} map with a 7 arcmin FWHM Gaussian and subtracted the resulting image from the original. We then stacked the central pixels of each galaxy to get co-added values in each of the \textit{Planck} bands. In addition, we degraded the ACT 148 and 220 GHz maps to match the \textit{Planck} beam, applied the same 7 arcmin FWHM filtering, and stacked the central pixels of galaxies in those images as well.

As was the case in Section \ref{sec:stack}, in all of these stacks there is an offset we needed to correct for since we are purposely avoiding positive contaminations in the maps. To do this we also made measurements at 54,962 random points on the sky that were restricted to the same contaminating-source cuts as our galaxies. These measurements allowed us to compute offset values needed to shift each band to a mean of 0, which we applied to our final measurements.

Finally, we computed our measurement errors by using the random point measurements \citep[since the proper noise covariance matrix is not provided, i.e.][]{PlanckCollaboration2015e}, corrected in two ways. First, because we account for the residual CMB primary signal in the $\chi^2$ calculations, discussed below, we removed the corresponding uncertainty term, taken to have a  covariance of 7.85 $\mu$K  as used in \citet{Spacek2016}. Second, there is an error introduced due to our offset corrections because they are made from a large, but finite, number of points. We then get the corrected error from
\begin{equation}
(\sigma_i \sigma_j)_{\text{corr}} = \sqrt{\frac{\sigma_i \sigma_j - \sigma_{\text{cov}}^2}{N_{\text{source}}} + \frac{\sigma_i \sigma_j}{N_{\text{random}}}},
\label{eq:newsig}
\end{equation}
where $\sigma_i \sigma_j$ is given by Equation (\ref{eq:sigmaact}) with $i$ and $j$ representing the various bands used, $\sigma_{\text{cov}} = 7.85 \, \mu$K is the minimum CMB covariance discussed above, $N_{\text{source}}$ is the number of sources used for the measurements (227 for low-$z$ and 529 for high-$z$), and $N_{\text{random}} = $ 54,962 is the number of random points used. This represents both the error due to detector noise in each band as well as the error due to contributions from foregrounds  on the sky. The majority of the variance at the highest frequencies is correlated between the bands and likely due to contributions from Galactic dust emission. However, unlike the primary CMB signal, the spectral shape of this foreground is similar to that of the dusty sources we are trying to constrain, and it cannot be removed by fitting it separately.  

In the same manner as the previous section, we modeled ACT 148 and 220 GHz contaminant source fluxes using a range of different source-count models (i.e. $\alpha_d,$ $\alpha_s,$ and $S_{\rm min}$), resulting again in 100,000 modeled contaminating source fluxes in each ACT band, $S_{148,\text{cont}}$ and $S_{220,\text{cont}}$. We also modeled what the contaminating signal would be in the \textit{Planck} bands and the ACT bands filtered to match \textit{Planck}.  For each modeled contaminating source, if it was chosen to be a synchrotron source we simply extrapolated the \textit{Planck}-based fluxes as
\begin{equation}
S_{\nu,\text{sync}} = S_{148,\text{cont}} \times \left( \frac{\nu}{\nu_{148}} \right)^{\alpha^{150}_{220}} \times C_\nu \times F ,
\label{eq:synchro}
\end{equation}
integrated over the relevant band filter curves, where $\alpha^{150}_{220}$ is the same used in the previous section, $C_\nu$ is a frequency-dependent factor involved in the conversion from Jy to $\mu$K arcmin$^2$, and $F = 0.021$ is the factor required to preserve the signal within a 1 arcmin radius aperture after applying the \textit{Planck} filtering we used.

In order to accurately describe  thermal dust emission  across the \textit{Planck} frequencies, we adopted a modified blackbody with a free emissivity index, $\beta$, and dust temperature, $T_\text{dust},$ often referred to as a gray-body \citep{PlanckMaps2015}.  This requires us to add another free parameter, the temperature of the contaminant dust, $T_\text{dust}$. 
This slope of each dusty source as a function of frequency is then
\begin{equation}
\left.\frac{d \ln S_\nu}{d \ln \nu}\right|_{\nu=185\, \text{GHz}} = 3 + \beta - x_{185} [1 - \exp(-x_{185})]^{-1} ,
\label{eq:dustslope}
\end{equation}
where $x_{185} \equiv (185$ GHz)$\times h/(kT) = (185/416) (1+z)/T_{20}$ and $T_{20}$ is the dust temperature in units of 20 K, and
we use the slope of the blackbody function at $\nu = 185$ GHz because it is halfway between our two ACT bands (148 and 220 GHz).  This can be related, in turn, to the power law index $\alpha^{150}_{220}$ as
\begin{equation}
\beta + 3 = \alpha^{150}_{220} + x_{185} [1 - \exp(-x_{185})]^{-1} .
\label{eq:beta}
\end{equation}
This then gives
\begin{equation}
\begin{split}
& S_{\nu,\text{dust}} = \,\, S_{148,\text{cont}} \times \left( \frac{\nu}{\nu_{148}} \right)^{\alpha^{150}_{220} + x_{185} [1 - \exp(-x_{185})]^{-1}}\\
&\times \frac{\exp[(\nu_{148}/416)(1+z)/T_{20}] - 1}{\exp[(\nu/416)(1+z)/T_{20}] - 1} \times C_\nu \times F,
\label{eq:sdust}
\end{split}
\end{equation}
integrated over the relevant band filter curves, where we vary $T_\text{dust}$ from 20 to 50 K in steps of 3 K.

With these expressions, we were able to compute $\chi^2$ values for each source-count model accounting for the \textit{Planck} measurements. This time, in addition to varying $f_{\text{cont}}$ and $S_{\text{SZ}}$, we also varied T$_\text{dust}$ (as discussed above) and a parameter $\Delta,$ which represents the offset due to the CMB primary signal, which we vary from -3 to 3 $\mu$K in steps of 0.1 $\mu$K. Computing $\chi^2$ now involved the original ACT terms plus the new \textit{Planck} terms, and it followed the same process as in Equation (\ref{eq:chi2matrixes}),
\begin{equation}
\chi^2(f_{\text{cont}},S_{\text{SZ}},T_\text{dust},\Delta) = \mathcal{B} \times \mathcal{A}^{-1} \times \mathcal{B}^T ,
\label{eq:chi2matrixplanck}
\end{equation}
where $\mathcal{B}$ is the signal array and $\mathcal{A}$ is the noise matrix containing the noise for each band plus the covariance terms between each band. We denote each element of the signal array $\mathcal{B}_i$, where $i$ runs over the two ACT bands (i.e. 148 and 220 GHz) and then every \textit{Planck}-filtered band (i.e. the \textit{Planck} bands at 857, 545, 353, and 217 GHz, plus the ACT bands at 220 and 148 GHz filtered to match the \textit{Planck} images), such that $\mathcal{B}_1 = f_{\text{cont}} \times \left<S_{148,\text{cont}}\right> + S_{\text{SZ}} - S_{148}$, $\mathcal{B}_2 = f_{\text{cont}} \times \left<S_{220,\text{cont}}\right> - S_{220}$, and $\mathcal{B}_{3-8} = f_{\text{cont}} \times \left<S_{3-8,\text{cont}}\right> + \Delta - S_{3-8}$. As before, $S_i$ represents the final values of our galaxy stacks for each band.  We similarly define the elements of the noise matrix as $\mathcal{A}_{i j} = \sigma_i \sigma_j$, where $i$ and $j$ run over all of the bands and $\sigma_i \sigma_j$ is given by Equation (\ref{eq:newsig}).

As in the previous section, we then converted the $\chi^2$ values to Gaussian probabilities by taking
\begin{equation}
\begin{split}
& P(S_{\text{SZ}}) = \\ & \sum_{f_{\text{cont}} \in [0,1],T_\text{dust},\Delta} \frac{\exp[-\chi^2(f_{\text{cont}},S_{\text{SZ}},T_\text{dust},\Delta)/2]}{\sum_{\text{all}}\exp[-\chi^2/2]},
\label{eq:probplanck}
\end{split}
\end{equation}
where the whole function is normalized to a total of 1, and each final SZ value contains the sum over the corresponding $T_\text{dust}$, $\Delta$, and fractions from 0 to 1. Since in this case there are eight terms contributing to $\chi^2$ and four fit parameters, this leaves us with four degrees of freedom.
Thus the minimum $\chi^2$ was not 0 in every case as it was above with just 2 measurements and 2 fit parameters, and so for each model we scale the final probabilities by $\exp(-\chi^2_{\text{min}}/2)$, where $\chi^2_{\text{min}}$ is the minimum $\chi^2$ value for that model.

\begin{figure*}[ht]
\centerline{\includegraphics[height=10cm]{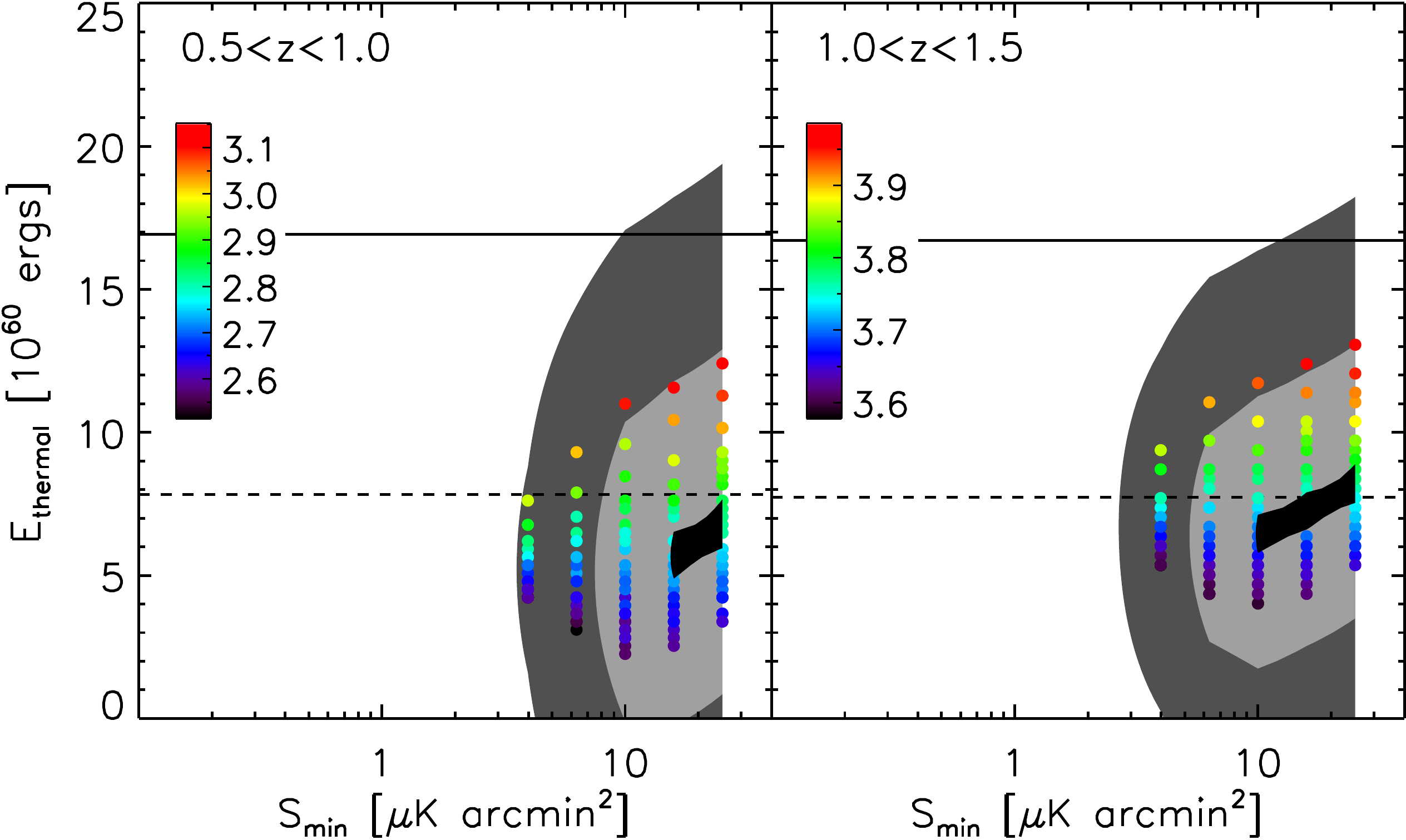}}
\caption{\small Plot of the contaminant-corrected $E_\text{therm}$ (see Equation (\ref{eq:newEthrm})) for different choices of $\alpha_{\rm dust}$, $\alpha_{\rm sync}$, and $S_{\rm min}$, incorporating the \textit{Planck} bands. Points are located at the peak $\chi^2$ probability for each model, and colored according to the minimum $\chi^2$ value for that model. Their general locations are still indicative of their $\alpha_d,$ $\alpha_s,$ and $S_{\rm min}$ values, as seen in Figure \ref{fig:s148corr}. The light and dark gray regions represent the complete span of $\pm1\sigma$ and $\pm2\sigma$, respectively, for all points, and the black regions represents the peak of the $\chi^2$ probability distribution, i.e. the most favorable models. The horizontal solid black lines represent the best estimates for $E_\text{grav}$, and the horizontal dashed black lines represent the $-1\sigma$ values for $E_\text{grav}$ (see Equation (\ref{eq:Egrav})).}
\label{fig:s148corrPlanck}
\end{figure*}

\begin{table*}[t]
\begin{center}
\resizebox{16cm}{!}{
\begin{tabular}{|c|c|c|c|c|c|}
\hline
Study     & N     & Type     & $z$ (mean)   & Mass ($M_\odot$)  & tSZ $Y (10^{-7}$ Mpc$^2$)  \\ \hline
\citet{Spacek2016} & 3394 & SPT & $0.5-1.0$ (0.72)  & $1.51 \times 10^{11}$ & $2.3^{+0.9}_{-0.7}$ \\              
\citet{Spacek2016} & 924 & SPT & $1.0-1.5$ (1.17) & $1.78 \times 10^{11}$ & $1.9^{+2.4}_{-2.0}$ \\
\citet{Spacek2016} & 937 & SPT+Planck & $0.5-1.0$ (0.72) & $1.51 \times 10^{11}$ & $2.2^{+0.9}_{-0.7}$ \\              
\citet{Spacek2016} & 240 & SPT+Planck & $1.0-1.5$ (1.17) & $1.78 \times 10^{11}$ & $1.7^{+2.2}_{-1.8}$ \\
Current & 1179 & ACT & $0.5-1.0$ (0.83) & $7.81 \times 10^{11}$ & $1.2_{-1.4}^{+1.4}$ \\
Current & 3274 & ACT & $1.0-1.5$ (1.20) & $10.1 \times 10^{11}$ & $1.9_{-1.0}^{+1.1}$ \\
Current & 227 & ACT+Planck & $0.5-1.0$ (0.83) & $6.93 \times 10^{11}$ & $1.5^{+1.6}_{-1.5}$ \\
Current & 529 & ACT+Planck & $1.0-1.5$ (1.21) & $9.68 \times 10^{11}$ & $1.9^{+1.3}_{-1.2}$ \\
\hline
\end{tabular} }
\end{center}
\caption{\small A comparison between \citet{Spacek2016} and the current work. $Y$ is the angularly integrated Compton-$y$ parameter given by Equation (\ref{eq:bigy}). Mass refers to stellar mass. \vspace{1mm}}
\label{tab:sptcompare}
\end{table*}

This again gave us a function $P(S_{\text{SZ}})$ for each combination of $\alpha_d,$ $\alpha_s,$ and $S_{\rm min}$, which we can convert to an energy $E_\text{therm}$. The peak of each $P(S_{\text{SZ}})$ curve is shown as the colored points in Figure \ref{fig:s148corrPlanck}, where the points are colored by the minimum $\chi^2$ value for each model. The 1$\sigma$ and 2$\sigma$ contours are created for each $S_{\rm min}$ by averaging $P(S_{\text{SZ}})$ across $\alpha_d$ and $\alpha_s$ and then dividing the final result by the single maximum value. 1$\sigma$ and 2$\sigma$ are again represented by the values 0.61 and 0.13, respectively, with peak probability values represented by $P(S_{\text{SZ}})>0.99$. These contours are shown in Figure \ref{fig:s148corrPlanck}, along with the $-1\sigma$ range for $E_\text{grav}$. From this figure we can see that, for low-$z$, including \textit{Planck} has slightly increased the estimated tSZ effect, though it now favors higher $S_{\text{min}}$ values. The high-$z$ result has not changed much besides an increased uncertainty due to fewer galaxies. For both redshift bins, the $\sigma$ values have increased due to the large decrease in the number of stacked galaxies because of \textit{Planck}'s much larger beam.

Finally, we average the probability distribution across $S_{\rm min}$, divided by the maximum value again, and get a final distribution as a function of only $E_\text{therm}$. The significance ($\sigma$) values of this curve are shown in the ``With \textit{Planck}'' part of Table \ref{tab:finvals}. At low-$z$, the significance of our tSZ detection has increased from $0.9\sigma$ to $1.0\sigma$, while at high-$z$ the tSZ detection significance decreases from $1.8\sigma$ to $1.6\sigma$. It is clear that the gain in sensitivity with \textit{Planck} has been limited by the decrease in the number of galaxies in each redshift bin due to the much larger beam size of \textit{Planck} compared to the ACT. To clearly show this, we followed the methods of Section \ref{sec:contaminant} (i.e. only ACT data were used), using this limited galaxy sample. The result is a $0.1\sigma$ tSZ detection at low-$z$ and a $0.5\sigma$ tSZ detection at high-$z$. It is apparent, then, that adding the \textit{Planck} data helps immensely with constraining the tSZ signal when the same number of galaxies are used, but since we have to limit our galaxy sample size so much to avoid contaminants in the \textit{Planck} beam, the loss of accuracy due to fewer measurements just about offsets the gain in accuracy given by the added \textit{Planck} data.

Alternatively, we can also characterize the total tSZ signal for our co-adds with the angularly integrated Compton-$y$ parameter, $Y$. While we cannot directly compare peak Compton-$y$ values with past measurements, as these are beam-dependent quantities, we can compare the angularly integrated $Y$ values between our results and past experiments. Using Equation (\ref{eq:newys}) at 148 GHz, this is
\begin{equation}
\begin{split}
Y &\equiv l_{\rm ang}^2 \int y(\bm{\theta}) d\bm{\theta} \\ &=  -3.2 \times 10^{-8} \, \text{Mpc}^2 
 \left(\frac{l_{\rm ang}}{\text{Gpc}}\right)^2 \frac{\int \Delta T_{148}(\bm{\theta}) d\bm{\theta}}{\text{$\mu$K arcmin$^2$}},
\label{eq:bigy}
\end{split}
\end{equation}
such that $Y = 2.7 \times 10^{-8}  \, \text{Mpc}^2 E_{\rm 60}$, where $E_{\rm 60}$ is $E_\text{therm}$ in units of $10^{60}$ erg. We can use this to compare the results in this paper with the similar work in\\ \citet{Spacek2016}, with a detailed comparison shown in Table\ \ref{tab:sptcompare}. Comparing the two results, we see a decrease in $Y$ at low-$z$ in this work compared to \citet{Spacek2016}, and similar $Y$ values at high-$z$, but we also see a significant increase in the average galaxy mass. This is contrary to the expected trend of higher $Y$ with higher mass seen in previous work \citep[e.g.][]{PlanckCollaboration2013,Greco2015,Ruan2015}.

\begin{figure}[t]
\centerline{\includegraphics[height=8cm]{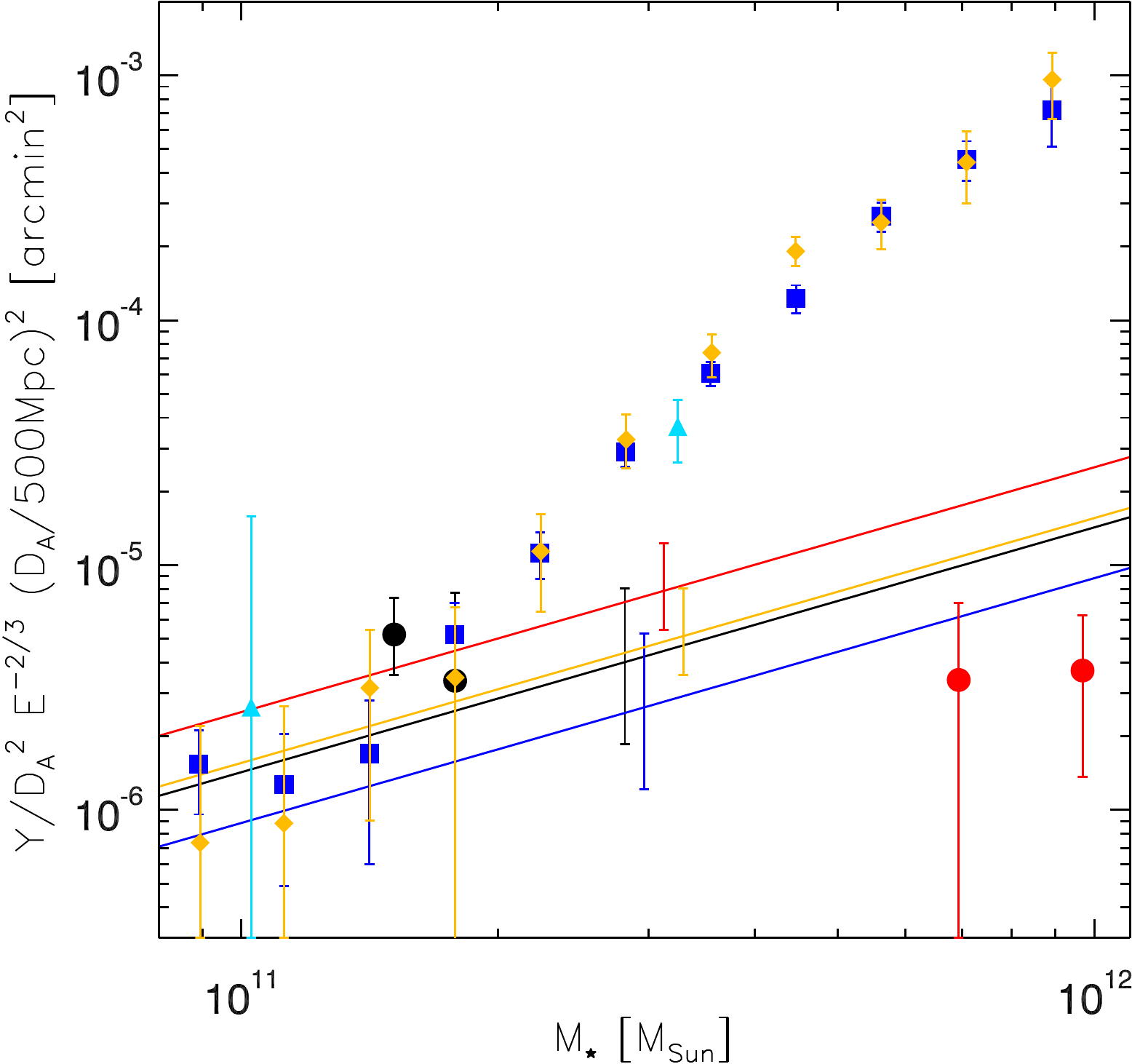}}
\caption{\small Plot of $\widetilde{Y}$ vs. stellar mass for \citet{Spacek2016} (black circles), the current work (red circles), \citet{PlanckCollaboration2013} (blue squares), \citet{Greco2015} (orange diamonds), and \citet{Ruan2015} (light blue triangles). Using Equations (\ref{eq:Egrav}), (\ref{eq:EAGN}), and (\ref{eq:bigy}), we can use our simple models to make estimates of $\widetilde{Y}$ vs. stellar mass. These model estimates are shown for gravitational heating only (black line for $z=0.8$, blue line for $z=1.2$) and gravitational plus AGN feedback heating (red line for $z=0.8$, orange line for $z=1.2$), with $\pm1\sigma$ errorbars. \vspace{3mm}}
\label{fig:ymassscale}
\end{figure}

We can compare with this previous work by defining $\widetilde{Y} \equiv \frac{Y}{l_{\rm ang}^2} \times E^{-2/3} \times \left(\frac{l_{\rm ang}}{500\ \text{Mpc}}\right)^2$, where $E(z)$ is the Hubble parameter, and this is shown in Figure \ref{fig:ymassscale}. The circles represent \citet{Spacek2016} (black) and this work (red), both using measurements of massive quiescent elliptical galaxies with average redshifts greater than 0.7 for the lower-mass values of either color and 1.1 for the higher-mass values of either color. The blue squares represent \citet{PlanckCollaboration2013} measurements of locally brightest galaxies, with redshifts less than $\approx$0.3. The orange diamonds represent \citet{Greco2015} measurements of locally brightest galaxies, with redshifts less than $\approx$0.3. The light blue triangles represent \citet{Ruan2015} measurements of locally brightest galaxies, with median redshifts of $\approx$0.5 and redshifts less than $\approx$0.8. Looking at this figure, we see that the results of \citet{Spacek2016} are roughly consistent with the previous tSZ measurements, while the results of this work are over $\approx$2 orders of magnitude smaller than previous tSZ measurements at the same mass. We note that the measurements of locally brightest galaxies from \citet{PlanckCollaboration2013}, \citet{Ruan2015}, and \citet{Greco2015} are of significantly lower redshifts than our galaxies, with the highest overall redshift being no more than $\approx$0.8 from \citet{Ruan2015} while our average redshifts range from 0.7 to 1.2. In addition, our selection criteria involve choosing quiescent elliptical galaxies and removing any galaxies in or around all detectable clusters, while the low-redshift locally brightest galaxies of the previous studies are more likely to be found in the centers of massive galaxy groups and clusters. Redshift alone cannot account for the $\approx2$ orders of magnitude difference in measurements, as is made clear by the black and red lines representing our models at $z = 0.8$ (without and with AGN feedback, respectively) and the blue and orange lines representing our models at $z = 1.2$ (without and with AGN feedback, respectively). However, the differences in redshift combined with the different galaxy selection methods suggest that the galaxies used in this paper could be from fundamentally different populations and environments.

With Equations (\ref{eq:Egrav}) and (\ref{eq:EAGN}) and the redshifts and masses from Table \ref{tab:meanvals}, we can also investigate theoretical thermal energies of the gas around elliptical galaxies due to both gravity and AGN feedback. Without \textit{Planck}, we estimate the gravitational heating energy to be $E_\text{therm,grav} = 17.0_{-9.1}^{+17.0} \times 10^{60}$ erg for our low-$z$ sample and $E_\text{therm,grav} = 16.7_{-9.0}^{+16.7}\times 10^{60}$ erg for our high-$z$ sample. We therefore measure excess non-gravitational energies of $E_\text{therm,feed,dat} = -12.5^{+17.7}_{-10.6} \times 10^{60}$ erg for low-$z$ and $E_\text{therm,feed,dat} = -9.7^{+17.1}_{-9.8} \times 10^{60}$ erg for high-$z$, both of these values consistent with zero detection. For completeness, we can plug these into the theoretical AGN feedback energy equation and solving for $\epsilon_{k}$, we get feedback efficiencies of $-4.8^{+6.8}_{-4.1}$\% for low-$z$ and $-3.8^{+6.7}_{-3.9}$\% for high-$z$.

With \textit{Planck}, we estimate the gravitational heating energy to be $E_\text{therm,grav} = 15.1_{-8.1}^{+15.1} \times 10^{60}$ erg for our low-$z$ sample and $E_\text{therm,grav} = 15.9_{-8.5}^{+15.9}\times 10^{60}$ erg for our high-$z$ sample. We therefore measure excess non-gravitational energies of $E_\text{therm,feed,dat} = -9.5^{+16.1}_{-10.0} \times 10^{60}$ erg for low-$z$, and $E_\text{therm,feed,dat} = -8.9^{+16.5}_{-9.7} \times 10^{60}$ erg for high-$z$, both consistent with 0. Plugging these into the theoretical AGN feedback energy equation and solving for $\epsilon_{k}$, we get feedback efficiencies of $-4.1^{+7.0}_{-4.4}$\% for low-$z$ and $-3.7^{+6.8}_{-4.0}$\% for high-$z$. These values are very uncertain and consistent with a detected AGN feedback signal of 0. They also do not rule out the suggested and measured $\approx$5\% \citep[e.g.][]{Scannapieco2005,Ruan2015,Spacek2016}. We also note that the feedback efficiencies stated in \citet{Spacek2016} are mistakenly off by a factor of 5, and they should be $7.3^{+6.6}_{-7.8}$\% for low-$z$ and $6.6^{+17.3}_{-15.3}$\% for high-$z$.


\section{Discussion}
\label{sec:discussion}

In this paper we have performed a stacking analysis of the tSZ signal around 4453 massive elliptical galaxies that are promising candidates for containing relic heating due to past episodes of AGN feedback. We split our selected galaxies into two redshifts bins, with 1179 galaxies in our ``low-$z$'' bin ($0.5 \leq z \leq 1.0$) and 3274 galaxies in our ``high-$z$'' bin ($1.0 \leq z \leq 1.5$). Our initial stacks were dominated by considerable contaminating emission which was much stronger at 220 GHz than at 148 GHz.
 Since dusty contaminant emission emits at both 148 and 220 GHz, as suggested by Figure \ref{fig:allfilters}, the large signals at 220 GHz, where the tSZ effect is expected to be negligible, indicate a corresponding large contaminant signal at 148 GHz, where the tSZ effect causes a decrement. We therefore performed an analysis of the contaminating signal by modeling potential undetected sources and running a $\chi^2$ probability test on the models. This revealed the underlying tSZ signal, with a 0.9$\sigma$ significance at low-$z$ and a 1.8$\sigma$ significance at high-$z$. Finally, in order to better constrain the stacked contaminating signal, we incorporated high-frequency \textit{Planck} measurements of a subset of 227 low-$z$ galaxies and 529 high-$z$ galaxies. These results indicated tSZ detections with a 1.0$\sigma$ significance at low-$z$ and a 1.6$\sigma$ significance at high-$z$. The values for each of these analyses are given in Table\ \ref{tab:finvals}.

The work done here is complementary to the work done in \citet{Spacek2016}, which stacked 4318 galaxies in a southern patch of sky using SPT data, while we stacked 4453 galaxies in the Stripe-82 equatorial band using ACT data. Both analyses used similar galaxy selection criteria, though that of \citet{Spacek2016} favored lower-mass, $0.5 \leq z \leq 1.0$ ``low-$z$'' galaxies while our selection favored $1.0 \leq z \leq 1.5$ ``high-$z$'' galaxies with higher stellar masses. Their most significant low-$z$ and high-$z$ tSZ detections were at $3.6\sigma$ and $0.9\sigma$ levels, respectively, while ours were at $1.0\sigma$ and $1.8\sigma$, respectively. A detailed comparison between the two studies can be seen in Table\ \ref{tab:sptcompare}, where we see similar tSZ $Y$ measurements in this work, although we use galaxies with higher masses. A plot comparing these results, as well as results from other previous galaxy tSZ measurements from \citet{PlanckCollaboration2013}, \citet{Greco2015}, and \citet{Ruan2015}, is shown in Figure \ref{fig:ymassscale}. These last three results appear to be significantly higher than the results of this paper. This may be due to several factors, including inherent differences in the measurements due to different galaxy populations. The previous studies focus on lower-redshift locally brightest galaxies, while this work looks at quiescent elliptical galaxies at significantly higher redshifts. We also perform extensive cuts to avoid clusters and dusty galaxies. There is the additional possibility that we are not completely accounting for and removing the contamination signal in this work despite our best efforts, though this seems unlikely to be the main reason for the discrepancy. Also shown in Figure \ref{fig:ymassscale} are lines representing our simple gravitational and AGN feedback heating models given by Equations (\ref{eq:Egrav}) and (\ref{eq:EAGN}). These simple models indicate that the types of galaxies and redshifts that we are looking at are expected to produce significantly lower tSZ measurements. The measurements presented here are also unique, with a review of the literature revealing no other similar measurements of the tSZ signal around such massive, high-redshift, quiescent elliptical galaxies. It therefore may not be completely appropriate to make direct comparisons between these measurements and measurements of less massive, lower-redshift locally brightest galaxies. While \citet{Spacek2016} estimate AGN feedback efficiencies of around $\approx$7\%, close to the suggested 5\% \citep[e.g.][]{Scannapieco2005,Ruan2015}, this work sees an AGN feedback heating signal consistent with 0, with efficiencies of $-4.1^{+7.0}_{-4.4}$\% for low-$z$ and $-3.7^{+6.8}_{-4.0}$\% for high-$z$. It is important to note, however, that we use simple, general models of gravitation and AGN feedback in this paper to estimate the corresponding energies, and that specific, detailed galaxy simulations are needed to draw more precise conclusions from these measurements.

tSZ measurements of galaxies and AGNs are likely to improve significantly in the near future. More data and an additional band at 277 GHz will be released from ACT observations \citep{Dunner2013}, while an upcoming full survey release of SPT data will include a 2500 deg$^2$ field using bands at 95, 150, and 220 GHz \citep{Schaffer2011}. These much larger fields with more bands will allow for a much larger set of galaxies to be co-added at more frequencies, vastly improving the signal-to-noise of the measurements and allowing for further constraints on contaminating signals. Separating out such contaminants will also become more effective with future surveys such as those to be carried out by the upgraded ACT telescope (Advanced ACTPol) and the proposed Cerro Chajnantor Atacama Telescope (CCAT).\footnote{http://www.ccatobservatory.org/} Another approach to  constraining AGN feedback is through deep measurements of smaller samples of galaxies, identified as the most interesting, using large radio telescopes. The Goddard IRAM Superconducting Two Millimeter Camera (GISMO)  and the New IRAM KIDs Array (NIKA) are powerful new instruments mounted on the Institute de Radioastronome Millimetrique (IRAM) 30 m telescope\footnote{http://www.iram-institute.org/EN/30-meter-telescope.php} that may prove useful for this purpose. Also promising is the National Radio Astronomy Observatories (NRAO) Green Bank Telescope (GBT), whose Continuum Backend operates at lower frequencies where the tSZ signal is roughly three times larger. Finally, although tSZ observations reveal the total thermal heating around galaxies, they must be complimented by theoretical models and simulations in order to best distinguish between heating due to gravitation, AGN feedback, and other effects. Observations can therefore be combined with tSZ simulations of the same types of objects with the same average parameters (e.g. mass, redshift, age) to produce weighted stacks that are adapted to be as sensitive as possible to the differences between AGN feedback models. The tSZ effect provides a promising tool for future measurements to improve our understanding of AGN feedback and galaxy evolution.

We would like to thank Arthur Kosowsky for helpful discussions.   This publication makes use of data products from the \textit{WISE}, which is a joint project of  UCLA, JPL, and Caltech funded by NASA. It also makes use of data produced from SDSS-III, which was funded by the Alfred P. Sloan Foundation, the National Science Foundation (NSF), the U.S. Department of Energy Office of Science, and the SDSS-III Participating Institutions.  Finally, we make use of data from the ACT project, which operates in the Parque Astron\'omico Atacama in northern Chile under the auspices of the Comisi\'on Nacional de Investigaci\'on Cient\'ifica y Tecnol\'ogica de Chile (CONICYT), and is funded by the NSF, Princeton University, U.\ Penn., and a Canada Foundation for Innovation award to UBC.  AS, ES, \& SC were supported by the NSF under grant AST14-07835.





\clearpage
\bibliographystyle{apj}
\small
\bibliography{references}

\begin{thebibliography}{}
\expandafter\ifx\csname natexlab\endcsname\relax\def\natexlab#1{#1}\fi

\bibitem[{{Alam} {et~al.}(2015){Alam}, {Albareti}, {Allende Prieto}, {Anders},
  {Anderson}, {Anderton}, {Andrews}, {Armengaud}, {Aubourg}, {Bailey}, \&
  et~al.}]{Alam2015}
{Alam}, S., {Albareti}, F.~D., {Allende Prieto}, C., {et~al.} 2015, \apjs, 219,
  12

\bibitem[{{Barger} {et~al.}(2005){Barger}, {Cowie}, {Mushotzky}, {Yang},
  {Wang}, {Steffen}, \& {Capak}}]{Barger2005}
{Barger}, A.~J., {Cowie}, L.~L., {Mushotzky}, R.~F., {et~al.} 2005, \aj, 129,
  578

\bibitem[{{Battaglia} {et~al.}(2010){Battaglia}, {Bond}, {Pfrommer}, {Sievers},
  \& {Sijacki}}]{Battaglia2010}
{Battaglia}, N., {Bond}, J.~R., {Pfrommer}, C., {Sievers}, J.~L., \& {Sijacki},
  D. 2010, \apj, 725, 91

\bibitem[{{Bauer} {et~al.}(2005){Bauer}, {Drory}, {Hill}, \&
  {Feulner}}]{Bauer2005}
{Bauer}, A.~E., {Drory}, N., {Hill}, G.~J., \& {Feulner}, G. 2005, \apjl, 621,
  L89

\bibitem[{{Best} \& {Heckman}(2012)}]{Best2012}
{Best}, P.~N., \& {Heckman}, T.~M. 2012, \mnras, 421, 1569

\bibitem[{{Best} {et~al.}(2005){Best}, {Kauffmann}, {Heckman}, {Brinchmann},
  {Charlot}, {Ivezi{\'c}}, \& {White}}]{Best2005}
{Best}, P.~N., {Kauffmann}, G., {Heckman}, T.~M., {et~al.} 2005, \mnras, 362,
  25

\bibitem[{{Birkinshaw}(1999)}]{Birkinshaw1999}
{Birkinshaw}, M. 1999, \physrep, 310, 97

\bibitem[{{B{\^i}rzan} {et~al.}(2004){B{\^i}rzan}, {Rafferty}, {McNamara},
  {Wise}, \& {Nulsen}}]{Birzan2004}
{B{\^i}rzan}, L., {Rafferty}, D.~A., {McNamara}, B.~R., {Wise}, M.~W., \&
  {Nulsen}, P.~E.~J. 2004, \apj, 607, 800

\bibitem[{{Bongiorno} {et~al.}(2016){Bongiorno}, {Schulze}, {Merloni},
  {Zamorani}, {Ilbert}, {La Franca}, {Peng}, {Piconcelli}, {Mainieri},
  {Silverman}, {Brusa}, {Fiore}, {Salvato}, \& {Scoville}}]{Bongiorno2016}
{Bongiorno}, A., {Schulze}, A., {Merloni}, A., {et~al.} 2016, \aap, 588, A78

\bibitem[{{Borguet} {et~al.}(2013){Borguet}, {Arav}, {Edmonds}, {Chamberlain},
  \& {Benn}}]{Borguet2013}
{Borguet}, B.~C.~J., {Arav}, N., {Edmonds}, D., {Chamberlain}, C., \& {Benn},
  C. 2013, \apj, 762, 49

\bibitem[{{Bower} {et~al.}(2006){Bower}, {Benson}, {Malbon}, {Helly}, {Frenk},
  {Baugh}, {Cole}, \& {Lacey}}]{Bower2006}
{Bower}, R.~G., {Benson}, A.~J., {Malbon}, R., {et~al.} 2006, \mnras, 370, 645

\bibitem[{{Brammer} {et~al.}(2008){Brammer}, {van Dokkum}, \&
  {Coppi}}]{Brammer2008}
{Brammer}, G.~B., {van Dokkum}, P.~G., \& {Coppi}, P. 2008, \apj, 686, 1503

\bibitem[{{Brinchmann} {et~al.}(2004){Brinchmann}, {Charlot}, {White},
  {Tremonti}, {Kauffmann}, {Heckman}, \& {Brinkmann}}]{Brinchmann2004}
{Brinchmann}, J., {Charlot}, S., {White}, S.~D.~M., {et~al.} 2004, \mnras, 351,
  1151

\bibitem[{{Bruzual} \& {Charlot}(2003)}]{Bruzual2003}
{Bruzual}, G., \& {Charlot}, S. 2003, \mnras, 344, 1000

\bibitem[{{Buchner} {et~al.}(2015){Buchner}, {Georgakakis}, {Nandra},
  {Brightman}, {Menzel}, {Liu}, {Hsu}, {Salvato}, {Rangel}, {Aird}, {Merloni},
  \& {Ross}}]{Buchner2015}
{Buchner}, J., {Georgakakis}, A., {Nandra}, K., {et~al.} 2015, \apj, 802, 89

\bibitem[{{Bundy} {et~al.}(2005){Bundy}, {Ellis}, \& {Conselice}}]{Bundy2005}
{Bundy}, K., {Ellis}, R.~S., \& {Conselice}, C.~J. 2005, \apj, 625, 621

\bibitem[{{Burns}(1990)}]{Burns1990}
{Burns}, J.~O. 1990, \aj, 99, 14

\bibitem[{{Cattaneo} {et~al.}(1999){Cattaneo}, {Haehnelt}, \&
  {Rees}}]{Cattaneo1999}
{Cattaneo}, A., {Haehnelt}, M.~G., \& {Rees}, M.~J. 1999, \mnras, 308, 77

\bibitem[{{Cen} \& {Safarzadeh}(2015)}]{Cen2015}
{Cen}, R., \& {Safarzadeh}, M. 2015, \apjl, 809, L32

\bibitem[{{Chamberlain} {et~al.}(2015){Chamberlain}, {Arav}, \&
  {Benn}}]{Chamberlain2015}
{Chamberlain}, C., {Arav}, N., \& {Benn}, C. 2015, \mnras, 450, 1085

\bibitem[{{Chartas} {et~al.}(2007){Chartas}, {Brandt}, {Gallagher}, \&
  {Proga}}]{Chartas2007}
{Chartas}, G., {Brandt}, W.~N., {Gallagher}, S.~C., \& {Proga}, D. 2007, \aj,
  133, 1849

\bibitem[{{Chatterjee} {et~al.}(2008){Chatterjee}, {Di Matteo}, {Kosowsky}, \&
  {Pelupessy}}]{Chatterjee2008}
{Chatterjee}, S., {Di Matteo}, T., {Kosowsky}, A., \& {Pelupessy}, I. 2008,
  \mnras, 390, 535

\bibitem[{{Chatterjee} {et~al.}(2010){Chatterjee}, {Ho}, {Newman}, \&
  {Kosowsky}}]{Chatterjee2010}
{Chatterjee}, S., {Ho}, S., {Newman}, J.~A., \& {Kosowsky}, A. 2010, \apj, 720,
  299

\bibitem[{{Chatterjee} \& {Kosowsky}(2007)}]{Chatterjee2007}
{Chatterjee}, S., \& {Kosowsky}, A. 2007, \apjl, 661, L113

\bibitem[{{Chen} {et~al.}(2009){Chen}, {Wild}, {Kauffmann}, {Blaizot}, {Davis},
  {Noeske}, {Wang}, \& {Willmer}}]{Chen2009}
{Chen}, Y.-M., {Wild}, V., {Kauffmann}, G., {et~al.} 2009, \mnras, 393, 406

\bibitem[{{Costa} {et~al.}(2014){Costa}, {Sijacki}, \& {Haehnelt}}]{Costa2014}
{Costa}, T., {Sijacki}, D., \& {Haehnelt}, M.~G. 2014, \mnras, 444, 2355

\bibitem[{{Cowie} \& {Barger}(2008)}]{Cowie2008}
{Cowie}, L.~L., \& {Barger}, A.~J. 2008, \apj, 686, 72

\bibitem[{{Cowie} {et~al.}(1996){Cowie}, {Songaila}, {Hu}, \&
  {Cohen}}]{Cowie1996}
{Cowie}, L.~L., {Songaila}, A., {Hu}, E.~M., \& {Cohen}, J.~G. 1996, \aj, 112,
  839

\bibitem[{{Crichton} {et~al.}(2016){Crichton}, {Gralla}, {Hall}, {Marriage},
  {Zakamska}, {Battaglia}, {Bond}, {Devlin}, {Hill}, {Hilton}, {Hincks},
  {Huffenberger}, {Hughes}, {Kosowsky}, {Moodley}, {Niemack}, {Page},
  {Partridge}, {Sievers}, {Sif{\'o}n}, {Staggs}, {Viero}, \&
  {Wollack}}]{Crichton2016}
{Crichton}, D., {Gralla}, M.~B., {Hall}, K., {et~al.} 2016, \mnras, 458, 1478

\bibitem[{{Daddi} {et~al.}(2004){Daddi}, {Cimatti}, {Renzini}, {Fontana},
  {Mignoli}, {Pozzetti}, {Tozzi}, \& {Zamorani}}]{Daddi2004}
{Daddi}, E., {Cimatti}, A., {Renzini}, A., {et~al.} 2004, \apj, 617, 746

\bibitem[{{Das} {et~al.}(2014){Das}, {Louis}, {Nolta}, {Addison},
  {Battistelli}, {Bond}, {Calabrese}, {Crichton}, {Devlin}, {Dicker},
  {Dunkley}, {D{\"u}nner}, {Fowler}, {Gralla}, {Hajian}, {Halpern},
  {Hasselfield}, {Hilton}, {Hincks}, {Hlozek}, {Huffenberger}, {Hughes},
  {Irwin}, {Kosowsky}, {Lupton}, {Marriage}, {Marsden}, {Menanteau}, {Moodley},
  {Niemack}, {Page}, {Partridge}, {Reese}, {Schmitt}, {Sehgal}, {Sherwin},
  {Sievers}, {Spergel}, {Staggs}, {Swetz}, {Switzer}, {Thornton}, {Trac}, \&
  {Wollack}}]{Das2014}
{Das}, S., {Louis}, T., {Nolta}, M.~R., {et~al.} 2014, \jcap, 4, 014

\bibitem[{{de Kool} {et~al.}(2001){de Kool}, {Arav}, {Becker}, {Gregg},
  {White}, {Laurent-Muehleisen}, {Price}, \& {Korista}}]{deKool2001}
{de Kool}, M., {Arav}, N., {Becker}, R.~H., {et~al.} 2001, \apj, 548, 609

\bibitem[{{Debuhr} {et~al.}(2010){Debuhr}, {Quataert}, {Ma}, \&
  {Hopkins}}]{Debuhr2010}
{Debuhr}, J., {Quataert}, E., {Ma}, C.-P., \& {Hopkins}, P. 2010, \mnras, 406,
  L55

\bibitem[{{Drory} \& {Alvarez}(2008)}]{Drory2008}
{Drory}, N., \& {Alvarez}, M. 2008, \apj, 680, 41

\bibitem[{{Dunn} {et~al.}(2010){Dunn}, {Bautista}, {Arav}, {Moe}, {Korista},
  {Costantini}, {Benn}, {Ellison}, \& {Edmonds}}]{Dunn2010}
{Dunn}, J.~P., {Bautista}, M., {Arav}, N., {et~al.} 2010, \apj, 709, 611

\bibitem[{{D{\"u}nner} {et~al.}(2013){D{\"u}nner}, {Hasselfield}, {Marriage},
  {Sievers}, {Acquaviva}, {Addison}, {Ade}, {Aguirre}, {Amiri}, {Appel},
  {Barrientos}, {Battistelli}, {Bond}, {Brown}, {Burger}, {Calabrese},
  {Chervenak}, {Das}, {Devlin}, {Dicker}, {Bertrand Doriese}, {Dunkley},
  {Essinger-Hileman}, {Fisher}, {Gralla}, {Fowler}, {Hajian}, {Halpern},
  {Hern{\'a}ndez-Monteagudo}, {Hilton}, {Hilton}, {Hincks}, {Hlozek},
  {Huffenberger}, {Hughes}, {Hughes}, {Infante}, {Irwin}, {Baptiste Juin},
  {Kaul}, {Klein}, {Kosowsky}, {Lau}, {Limon}, {Lin}, {Louis}, {Lupton},
  {Marsden}, {Martocci}, {Mauskopf}, {Menanteau}, {Moodley}, {Moseley},
  {Netterfield}, {Niemack}, {Nolta}, {Page}, {Parker}, {Partridge}, {Quintana},
  {Reid}, {Sehgal}, {Sherwin}, {Spergel}, {Staggs}, {Swetz}, {Switzer},
  {Thornton}, {Trac}, {Tucker}, {Warne}, {Wilson}, {Wollack}, \&
  {Zhao}}]{Dunner2013}
{D{\"u}nner}, R., {Hasselfield}, M., {Marriage}, T.~A., {et~al.} 2013, \apj,
  762, 10

\bibitem[{{Eisenhardt} {et~al.}(2012){Eisenhardt}, {Wu}, {Tsai}, {Assef},
  {Benford}, {Blain}, {Bridge}, {Condon}, {Cushing}, {Cutri}, {Evans},
  {Gelino}, {Griffith}, {Grillmair}, {Jarrett}, {Lonsdale}, {Masci}, {Mason},
  {Petty}, {Sayers}, {Stanford}, {Stern}, {Wright}, \& {Yan}}]{Eisenhardt2012}
{Eisenhardt}, P.~R.~M., {Wu}, J., {Tsai}, C.-W., {et~al.} 2012, \apj, 755, 173

\bibitem[{{Fabian}(2012)}]{Fabian2012}
{Fabian}, A.~C. 2012, \araa, 50, 455

\bibitem[{{Farrah} {et~al.}(2012){Farrah}, {Urrutia}, {Lacy}, {Efstathiou},
  {Afonso}, {Coppin}, {Hall}, {Lonsdale}, {Jarrett}, {Bridge}, {Borys}, \&
  {Petty}}]{Farrah2012}
{Farrah}, D., {Urrutia}, T., {Lacy}, M., {et~al.} 2012, \apj, 745, 178

\bibitem[{{Faucher-Gigu{\`e}re} \& {Quataert}(2012)}]{Faucher2012}
{Faucher-Gigu{\`e}re}, C.-A., \& {Quataert}, E. 2012, \mnras, 425, 605

\bibitem[{{Feldmann} \& {Mayer}(2015)}]{Feldmann2015}
{Feldmann}, R., \& {Mayer}, L. 2015, \mnras, 446, 1939

\bibitem[{{Ferrarese}(2002)}]{Ferrarese2002}
{Ferrarese}, L. 2002, \apj, 578, 90

\bibitem[{{Feruglio} {et~al.}(2010){Feruglio}, {Maiolino}, {Piconcelli},
  {Menci}, {Aussel}, {Lamastra}, \& {Fiore}}]{Feruglio2010}
{Feruglio}, C., {Maiolino}, R., {Piconcelli}, E., {et~al.} 2010, \aap, 518,
  L155

\bibitem[{{Feulner} {et~al.}(2005){Feulner}, {Gabasch}, {Salvato}, {Drory},
  {Hopp}, \& {Bender}}]{Feulner2005}
{Feulner}, G., {Gabasch}, A., {Salvato}, M., {et~al.} 2005, \apjl, 633, L9

\bibitem[{{Fontanot} {et~al.}(2009){Fontanot}, {De Lucia}, {Monaco},
  {Somerville}, \& {Santini}}]{Fontanot2009}
{Fontanot}, F., {De Lucia}, G., {Monaco}, P., {Somerville}, R.~S., \&
  {Santini}, P. 2009, \mnras, 397, 1776

\bibitem[{{Ganguly} \& {Brotherton}(2008)}]{Ganguly2008}
{Ganguly}, R., \& {Brotherton}, M.~S. 2008, \apj, 672, 102

\bibitem[{{Gralla} {et~al.}(2014){Gralla}, {Crichton}, {Marriage}, {Mo},
  {Aguirre}, {Addison}, {Asboth}, {Battaglia}, {Bock}, {Bond}, {Devlin},
  {D{\"u}nner}, {Hajian}, {Halpern}, {Hilton}, {Hincks}, {Hlozek},
  {Huffenberger}, {Hughes}, {Ivison}, {Kosowsky}, {Lin}, {Marsden},
  {Menanteau}, {Moodley}, {Morales}, {Niemack}, {Oliver}, {Page}, {Partridge},
  {Reese}, {Rojas}, {Sehgal}, {Sievers}, {Sif{\'o}n}, {Spergel}, {Staggs},
  {Switzer}, {Viero}, {Wollack}, \& {Zemcov}}]{Gralla2014}
{Gralla}, M.~B., {Crichton}, D., {Marriage}, T.~A., {et~al.} 2014, \mnras, 445,
  460

\bibitem[{{Greco} {et~al.}(2015){Greco}, {Hill}, {Spergel}, \&
  {Battaglia}}]{Greco2015}
{Greco}, J.~P., {Hill}, J.~C., {Spergel}, D.~N., \& {Battaglia}, N. 2015, \apj,
  808, 151

\bibitem[{{Greene} {et~al.}(2012){Greene}, {Zakamska}, \& {Smith}}]{Green2012}
{Greene}, J.~E., {Zakamska}, N.~L., \& {Smith}, P.~S. 2012, \apj, 746, 86

\bibitem[{{Grogin} {et~al.}(2011){Grogin}, {Kocevski}, {Faber}, {Ferguson},
  {Koekemoer}, {Riess}, {Acquaviva}, {Alexander}, {Almaini}, {Ashby}, {Barden},
  {Bell}, {Bournaud}, {Brown}, {Caputi}, {Casertano}, {Cassata}, {Castellano},
  {Challis}, {Chary}, {Cheung}, {Cirasuolo}, {Conselice}, {Roshan Cooray},
  {Croton}, {Daddi}, {Dahlen}, {Dav{\'e}}, {de Mello}, {Dekel}, {Dickinson},
  {Dolch}, {Donley}, {Dunlop}, {Dutton}, {Elbaz}, {Fazio}, {Filippenko},
  {Finkelstein}, {Fontana}, {Gardner}, {Garnavich}, {Gawiser}, {Giavalisco},
  {Grazian}, {Guo}, {Hathi}, {H{\"a}ussler}, {Hopkins}, {Huang}, {Huang},
  {Jha}, {Kartaltepe}, {Kirshner}, {Koo}, {Lai}, {Lee}, {Li}, {Lotz}, {Lucas},
  {Madau}, {McCarthy}, {McGrath}, {McIntosh}, {McLure}, {Mobasher},
  {Moustakas}, {Mozena}, {Nandra}, {Newman}, {Niemi}, {Noeske}, {Papovich},
  {Pentericci}, {Pope}, {Primack}, {Rajan}, {Ravindranath}, {Reddy}, {Renzini},
  {Rix}, {Robaina}, {Rodney}, {Rosario}, {Rosati}, {Salimbeni}, {Scarlata},
  {Siana}, {Simard}, {Smidt}, {Somerville}, {Spinrad}, {Straughn}, {Strolger},
  {Telford}, {Teplitz}, {Trump}, {van der Wel}, {Villforth}, {Wechsler},
  {Weiner}, {Wiklind}, {Wild}, {Wilson}, {Wuyts}, {Yan}, \& {Yun}}]{Grogin2011}
{Grogin}, N.~A., {Kocevski}, D.~D., {Faber}, S.~M., {et~al.} 2011, \apjs, 197,
  35

\bibitem[{{Haehnelt} \& {Tegmark}(1996)}]{Haehnelt1996}
{Haehnelt}, M.~G., \& {Tegmark}, M. 1996, \mnras, 279, 545

\bibitem[{{Hamann} {et~al.}(2001){Hamann}, {Barlow}, {Chaffee}, {Foltz}, \&
  {Weymann}}]{Hamann2001}
{Hamann}, F.~W., {Barlow}, T.~A., {Chaffee}, F.~C., {Foltz}, C.~B., \&
  {Weymann}, R.~J. 2001, \apj, 550, 142

\bibitem[{{Hand} {et~al.}(2011){Hand}, {Appel}, {Battaglia}, {Bond}, {Das},
  {Devlin}, {Dunkley}, {D{\"u}nner}, {Essinger-Hileman}, {Fowler}, {Hajian},
  {Halpern}, {Hasselfield}, {Hilton}, {Hincks}, {Hlozek}, {Hughes}, {Irwin},
  {Klein}, {Kosowsky}, {Lin}, {Marriage}, {Marsden}, {McLaren}, {Menanteau},
  {Moodley}, {Niemack}, {Nolta}, {Page}, {Parker}, {Partridge}, {Plimpton},
  {Reese}, {Rojas}, {Sehgal}, {Sherwin}, {Sievers}, {Spergel}, {Staggs},
  {Swetz}, {Switzer}, {Thornton}, {Trac}, {Visnjic}, \& {Wollack}}]{Hand2011}
{Hand}, N., {Appel}, J.~W., {Battaglia}, N., {et~al.} 2011, \apj, 736, 39

\bibitem[{{Hasselfield} {et~al.}(2013){Hasselfield}, {Hilton}, {Marriage},
  {Addison}, {Barrientos}, {Battaglia}, {Battistelli}, {Bond}, {Crichton},
  {Das}, {Devlin}, {Dicker}, {Dunkley}, {D{\"u}nner}, {Fowler}, {Gralla},
  {Hajian}, {Halpern}, {Hincks}, {Hlozek}, {Hughes}, {Infante}, {Irwin},
  {Kosowsky}, {Marsden}, {Menanteau}, {Moodley}, {Niemack}, {Nolta}, {Page},
  {Partridge}, {Reese}, {Schmitt}, {Sehgal}, {Sherwin}, {Sievers}, {Sif{\'o}n},
  {Spergel}, {Staggs}, {Swetz}, {Switzer}, {Thornton}, {Trac}, \&
  {Wollack}}]{Hasselfield2013}
{Hasselfield}, M., {Hilton}, M., {Marriage}, T.~A., {et~al.} 2013, \jcap, 7,
  008

\bibitem[{{Helou} \& {Walker}(1988)}]{IRAS1988}
{Helou}, G., \& {Walker}, D.~W., eds. 1988, {Infrared astronomical satellite
  (IRAS) catalogs and atlases. Volume 7: The small scale structure catalog},
  Vol.~7, 1--265

\bibitem[{{Hewett} \& {Foltz}(2003)}]{Hewett2003}
{Hewett}, P.~C., \& {Foltz}, C.~B. 2003, \aj, 125, 1784

\bibitem[{{Hirschmann} {et~al.}(2014){Hirschmann}, {Dolag}, {Saro}, {Bachmann},
  {Borgani}, \& {Burkert}}]{Hirschmann2014}
{Hirschmann}, M., {Dolag}, K., {Saro}, A., {et~al.} 2014, \mnras, 442, 2304

\bibitem[{{Hirschmann} {et~al.}(2012){Hirschmann}, {Somerville}, {Naab}, \&
  {Burkert}}]{Hirschmann2012}
{Hirschmann}, M., {Somerville}, R.~S., {Naab}, T., \& {Burkert}, A. 2012,
  \mnras, 426, 237

\bibitem[{{Ishihara} {et~al.}(2010){Ishihara}, {Onaka}, {Kataza}, {Salama},
  {Alfageme}, {Cassatella}, {Cox}, {Garc{\'{\i}}a-Lario}, {Stephenson},
  {Cohen}, {Fujishiro}, {Fujiwara}, {Hasegawa}, {Ita}, {Kim}, {Matsuhara},
  {Murakami}, {M{\"u}ller}, {Nakagawa}, {Ohyama}, {Oyabu}, {Pyo}, {Sakon},
  {Shibai}, {Takita}, {Tanab{\'e}}, {Uemizu}, {Ueno}, {Usui}, {Wada},
  {Watarai}, {Yamamura}, \& {Yamauchi}}]{Ishihara2010}
{Ishihara}, D., {Onaka}, T., {Kataza}, H., {et~al.} 2010, \aap, 514, A1

\bibitem[{{Kang} {et~al.}(2016){Kang}, {Kim}, {Lim}, {Chung}, \&
  {Lee}}]{Kang2016}
{Kang}, Y., {Kim}, Y.-L., {Lim}, D., {Chung}, C., \& {Lee}, Y.-W. 2016, \apjs,
  223, 7

\bibitem[{{Kauffmann} \& {Haehnelt}(2000)}]{Kauffmann2000}
{Kauffmann}, G., \& {Haehnelt}, M. 2000, \mnras, 311, 576

\bibitem[{{Keller} {et~al.}(2016){Keller}, {Wadsley}, \&
  {Couchman}}]{Keller2016}
{Keller}, B.~W., {Wadsley}, J., \& {Couchman}, H.~M.~P. 2016, \mnras,
  arXiv:1604.08244

\bibitem[{{Knigge} {et~al.}(2008){Knigge}, {Scaringi}, {Goad}, \&
  {Cottis}}]{Knigge2008}
{Knigge}, C., {Scaringi}, S., {Goad}, M.~R., \& {Cottis}, C.~E. 2008, \mnras,
  386, 1426

\bibitem[{{Kodama} {et~al.}(2004){Kodama}, {Yamada}, {Akiyama}, {Aoki}, {Doi},
  {Furusawa}, {Fuse}, {Imanishi}, {Ishida}, {Iye}, {Kajisawa}, {Karoji},
  {Kobayashi}, {Komiyama}, {Kosugi}, {Maeda}, {Miyazaki}, {Mizumoto},
  {Morokuma}, {Nakata}, {Noumaru}, {Ogasawara}, {Ouchi}, {Sasaki}, {Sekiguchi},
  {Shimasaku}, {Simpson}, {Takata}, {Tanaka}, {Ueda}, {Yasuda}, \&
  {Yoshida}}]{Kodama2004}
{Kodama}, T., {Yamada}, T., {Akiyama}, M., {et~al.} 2004, \mnras, 350, 1005

\bibitem[{{Koekemoer} {et~al.}(2011){Koekemoer}, {Faber}, {Ferguson}, {Grogin},
  {Kocevski}, {Koo}, {Lai}, {Lotz}, {Lucas}, {McGrath}, {Ogaz}, {Rajan},
  {Riess}, {Rodney}, {Strolger}, {Casertano}, {Castellano}, {Dahlen},
  {Dickinson}, {Dolch}, {Fontana}, {Giavalisco}, {Grazian}, {Guo}, {Hathi},
  {Huang}, {van der Wel}, {Yan}, {Acquaviva}, {Alexander}, {Almaini}, {Ashby},
  {Barden}, {Bell}, {Bournaud}, {Brown}, {Caputi}, {Cassata}, {Challis},
  {Chary}, {Cheung}, {Cirasuolo}, {Conselice}, {Roshan Cooray}, {Croton},
  {Daddi}, {Dav{\'e}}, {de Mello}, {de Ravel}, {Dekel}, {Donley}, {Dunlop},
  {Dutton}, {Elbaz}, {Fazio}, {Filippenko}, {Finkelstein}, {Frazer}, {Gardner},
  {Garnavich}, {Gawiser}, {Gruetzbauch}, {Hartley}, {H{\"a}ussler},
  {Herrington}, {Hopkins}, {Huang}, {Jha}, {Johnson}, {Kartaltepe},
  {Khostovan}, {Kirshner}, {Lani}, {Lee}, {Li}, {Madau}, {McCarthy},
  {McIntosh}, {McLure}, {McPartland}, {Mobasher}, {Moreira}, {Mortlock},
  {Moustakas}, {Mozena}, {Nandra}, {Newman}, {Nielsen}, {Niemi}, {Noeske},
  {Papovich}, {Pentericci}, {Pope}, {Primack}, {Ravindranath}, {Reddy},
  {Renzini}, {Rix}, {Robaina}, {Rosario}, {Rosati}, {Salimbeni}, {Scarlata},
  {Siana}, {Simard}, {Smidt}, {Snyder}, {Somerville}, {Spinrad}, {Straughn},
  {Telford}, {Teplitz}, {Trump}, {Vargas}, {Villforth}, {Wagner}, {Wandro},
  {Wechsler}, {Weiner}, {Wiklind}, {Wild}, {Wilson}, {Wuyts}, \&
  {Yun}}]{Koekemoer2011}
{Koekemoer}, A.~M., {Faber}, S.~M., {Ferguson}, H.~C., {et~al.} 2011, \apjs,
  197, 36

\bibitem[{{LaMassa} {et~al.}(2013){LaMassa}, {Urry}, {Cappelluti}, {Civano},
  {Ranalli}, {Glikman}, {Treister}, {Richards}, {Ballantyne}, {Stern},
  {Comastri}, {Cardamone}, {Schawinski}, {B{\"o}hringer}, {Chon}, {Murray},
  {Green}, \& {Nandra}}]{LaMassa2013}
{LaMassa}, S.~M., {Urry}, C.~M., {Cappelluti}, N., {et~al.} 2013, \mnras, 436,
  3581

\bibitem[{{Lanz} {et~al.}(2016){Lanz}, {Ogle}, {Alatalo}, \&
  {Appleton}}]{Lanz2016}
{Lanz}, L., {Ogle}, P.~M., {Alatalo}, K., \& {Appleton}, P.~N. 2016, \apj, 826,
  29

\bibitem[{{Lapi} {et~al.}(2003){Lapi}, {Cavaliere}, \& {De Zotti}}]{Lapi2003}
{Lapi}, A., {Cavaliere}, A., \& {De Zotti}, G. 2003, \apjl, 597, L93

\bibitem[{{Lapi} {et~al.}(2014){Lapi}, {Raimundo}, {Aversa}, {Cai}, {Negrello},
  {Celotti}, {De Zotti}, \& {Danese}}]{Lapi2014}
{Lapi}, A., {Raimundo}, S., {Aversa}, R., {et~al.} 2014, \apj, 782, 69

\bibitem[{{Marconi} \& {Hunt}(2003)}]{Marconi2003}
{Marconi}, A., \& {Hunt}, L.~K. 2003, \apjl, 589, L21

\bibitem[{{Marriage} {et~al.}(2011){Marriage}, {Acquaviva}, {Ade}, {Aguirre},
  {Amiri}, {Appel}, {Barrientos}, {Battistelli}, {Bond}, {Brown}, {Burger},
  {Chervenak}, {Das}, {Devlin}, {Dicker}, {Bertrand Doriese}, {Dunkley},
  {D{\"u}nner}, {Essinger-Hileman}, {Fisher}, {Fowler}, {Hajian}, {Halpern},
  {Hasselfield}, {Hern{\'a}ndez-Monteagudo}, {Hilton}, {Hilton}, {Hincks},
  {Hlozek}, {Huffenberger}, {Handel Hughes}, {Hughes}, {Infante}, {Irwin},
  {Baptiste Juin}, {Kaul}, {Klein}, {Kosowsky}, {Lau}, {Limon}, {Lin},
  {Lupton}, {Marsden}, {Martocci}, {Mauskopf}, {Menanteau}, {Moodley},
  {Moseley}, {Netterfield}, {Niemack}, {Nolta}, {Page}, {Parker}, {Partridge},
  {Quintana}, {Reese}, {Reid}, {Sehgal}, {Sherwin}, {Sievers}, {Spergel},
  {Staggs}, {Swetz}, {Switzer}, {Thornton}, {Trac}, {Tucker}, {Warne},
  {Wilson}, {Wollack}, \& {Zhao}}]{Marriage2011}
{Marriage}, T.~A., {Acquaviva}, V., {Ade}, P.~A.~R., {et~al.} 2011, \apj, 737,
  61

\bibitem[{{McNamara} {et~al.}(2005){McNamara}, {Nulsen}, {Wise}, {Rafferty},
  {Carilli}, {Sarazin}, \& {Blanton}}]{McNamara2005}
{McNamara}, B.~R., {Nulsen}, P.~E.~J., {Wise}, M.~W., {et~al.} 2005, \nat, 433,
  45

\bibitem[{{Menci}(2006)}]{Menci2006}
{Menci}, N. 2006, in ESA Special Publication, Vol. 604, The X-ray Universe
  2005, ed. A.~{Wilson}, 557

\bibitem[{{Merloni}(2004)}]{Merloni2004}
{Merloni}, A. 2004, \mnras, 353, 1035

\bibitem[{{Merloni} \& {Heinz}(2008)}]{Merloni2008}
{Merloni}, A., \& {Heinz}, S. 2008, \mnras, 388, 1011

\bibitem[{{Mocanu} {et~al.}(2013){Mocanu}, {Crawford}, {Vieira}, {Aird},
  {Aravena}, {Austermann}, {Benson}, {B{\'e}thermin}, {Bleem}, {Bothwell},
  {Carlstrom}, {Chang}, {Chapman}, {Cho}, {Crites}, {de Haan}, {Dobbs},
  {Everett}, {George}, {Halverson}, {Harrington}, {Hezaveh}, {Holder},
  {Holzapfel}, {Hoover}, {Hrubes}, {Keisler}, {Knox}, {Lee}, {Leitch},
  {Lueker}, {Luong-Van}, {Marrone}, {McMahon}, {Mehl}, {Meyer}, {Mohr},
  {Montroy}, {Natoli}, {Padin}, {Plagge}, {Pryke}, {Rest}, {Reichardt}, {Ruhl},
  {Sayre}, {Schaffer}, {Shirokoff}, {Spieler}, {Spilker}, {Stalder},
  {Staniszewski}, {Stark}, {Story}, {Switzer}, {Vanderlinde}, \&
  {Williamson}}]{Mocanu2013}
{Mocanu}, L.~M., {Crawford}, T.~M., {Vieira}, J.~D., {et~al.} 2013, \apj, 779,
  61

\bibitem[{{Mocz} {et~al.}(2013){Mocz}, {Fabian}, \& {Blundell}}]{Mocz2013}
{Mocz}, P., {Fabian}, A.~C., \& {Blundell}, K.~M. 2013, \mnras, 432, 3381

\bibitem[{{Moe} {et~al.}(2009){Moe}, {Arav}, {Bautista}, \&
  {Korista}}]{Moe2009}
{Moe}, M., {Arav}, N., {Bautista}, M.~A., \& {Korista}, K.~T. 2009, \apj, 706,
  525

\bibitem[{{Natarajan} \& {Sigurdsson}(1999)}]{Natarajan1999}
{Natarajan}, P., \& {Sigurdsson}, S. 1999, \mnras, 302, 288

\bibitem[{{Neistein} {et~al.}(2006){Neistein}, {van den Bosch}, \&
  {Dekel}}]{Neistein2006}
{Neistein}, E., {van den Bosch}, F.~C., \& {Dekel}, A. 2006, \mnras, 372, 933

\bibitem[{{Newton} \& {Kay}(2013)}]{Newton2013}
{Newton}, R.~D.~A., \& {Kay}, S.~T. 2013, \mnras, 434, 3606

\bibitem[{{Noeske} {et~al.}(2007){Noeske}, {Faber}, {Weiner}, {Koo}, {Primack},
  {Dekel}, {Papovich}, {Conselice}, {Le Floc'h}, {Rieke}, {Coil}, {Lotz},
  {Somerville}, \& {Bundy}}]{Noeske2007}
{Noeske}, K.~G., {Faber}, S.~M., {Weiner}, B.~J., {et~al.} 2007, \apjl, 660,
  L47

\bibitem[{{Oke} \& {Gunn}(1983)}]{Oke1983}
{Oke}, J.~B., \& {Gunn}, J.~E. 1983, \apj, 266, 713

\bibitem[{{Ostriker} {et~al.}(2010){Ostriker}, {Choi}, {Ciotti}, {Novak}, \&
  {Proga}}]{Ostriker2010}
{Ostriker}, J.~P., {Choi}, E., {Ciotti}, L., {Novak}, G.~S., \& {Proga}, D.
  2010, \apj, 722, 642

\bibitem[{{Page} {et~al.}(2012){Page}, {Symeonidis}, {Vieira}, {Altieri},
  {Amblard}, {Arumugam}, {Aussel}, {Babbedge}, {Blain}, {Bock}, {Boselli},
  {Buat}, {Castro-Rodr{\'{\i}}guez}, {Cava}, {Chanial}, {Clements}, {Conley},
  {Conversi}, {Cooray}, {Dowell}, {Dubois}, {Dunlop}, {Dwek}, {Dye}, {Eales},
  {Elbaz}, {Farrah}, {Fox}, {Franceschini}, {Gear}, {Glenn}, {Griffin},
  {Halpern}, {Hatziminaoglou}, {Ibar}, {Isaak}, {Ivison}, {Lagache},
  {Levenson}, {Lu}, {Madden}, {Maffei}, {Mainetti}, {Marchetti}, {Nguyen},
  {O'Halloran}, {Oliver}, {Omont}, {Panuzzo}, {Papageorgiou}, {Pearson},
  {P{\'e}rez-Fournon}, {Pohlen}, {Rawlings}, {Rigopoulou}, {Riguccini},
  {Rizzo}, {Rodighiero}, {Roseboom}, {Rowan-Robinson}, {Portal}, {Schulz},
  {Scott}, {Seymour}, {Shupe}, {Smith}, {Stevens}, {Trichas}, {Tugwell},
  {Vaccari}, {Valtchanov}, {Viero}, {Vigroux}, {Wang}, {Ward}, {Wright}, {Xu},
  \& {Zemcov}}]{Page2012}
{Page}, M.~J., {Symeonidis}, M., {Vieira}, J.~D., {et~al.} 2012, \nat, 485, 213

\bibitem[{{Papovich} {et~al.}(2006){Papovich}, {Moustakas}, {Dickinson}, {Le
  Floc'h}, {Rieke}, {Daddi}, {Alexander}, {Bauer}, {Brandt}, {Dahlen}, {Egami},
  {Eisenhardt}, {Elbaz}, {Ferguson}, {Giavalisco}, {Lucas}, {Mobasher},
  {P{\'e}rez-Gonz{\'a}lez}, {Stutz}, {Rieke}, \& {Yan}}]{Papovich2006}
{Papovich}, C., {Moustakas}, L.~A., {Dickinson}, M., {et~al.} 2006, \apj, 640,
  92

\bibitem[{{Pei}(1995)}]{Pei1995}
{Pei}, Y.~C. 1995, \apj, 438, 623

\bibitem[{{Peng} {et~al.}(2002){Peng}, {Ho}, {Impey}, \& {Rix}}]{Peng2002}
{Peng}, C.~Y., {Ho}, L.~C., {Impey}, C.~D., \& {Rix}, H.-W. 2002, \aj, 124, 266

\bibitem[{{Piffaretti} {et~al.}(2011){Piffaretti}, {Arnaud}, {Pratt},
  {Pointecouteau}, \& {Melin}}]{Piffaretti2011}
{Piffaretti}, R., {Arnaud}, M., {Pratt}, G.~W., {Pointecouteau}, E., \&
  {Melin}, J.-B. 2011, \aap, 534, A109

\bibitem[{{Pipino} {et~al.}(2009){Pipino}, {Silk}, \& {Matteucci}}]{Pipino2009}
{Pipino}, A., {Silk}, J., \& {Matteucci}, F. 2009, \mnras, 392, 475

\bibitem[{{Planck Collaboration} {et~al.}(2013){Planck Collaboration}, {Ade},
  {Aghanim}, {Arnaud}, {Ashdown}, {Atrio-Barandela}, {Aumont}, {Baccigalupi},
  {Balbi}, {Banday}, \& et~al.}]{PlanckCollaboration2013}
{Planck Collaboration}, {Ade}, P.~A.~R., {Aghanim}, N., {et~al.} 2013, \aap,
  557, A52

\bibitem[{{Planck Collaboration} {et~al.}(2014){Planck Collaboration}, {Ade},
  {Aghanim}, {Arg{\"u}eso}, {Armitage-Caplan}, {Arnaud}, {Ashdown},
  {Atrio-Barandela}, {Aumont}, {Baccigalupi}, \&
  et~al.}]{PlanckCollaboration2014}
---. 2014, \aap, 571, A28

\bibitem[{{Planck Collaboration} {et~al.}(2015{\natexlab{a}}){Planck
  Collaboration}, {Adam}, {Ade}, {Aghanim}, {Arnaud}, {Ashdown}, {Aumont},
  {Baccigalupi}, {Banday}, {Barreiro}, \& et~al.}]{PlanckCollaboration2015d}
{Planck Collaboration}, {Adam}, R., {Ade}, P.~A.~R., {et~al.}
  2015{\natexlab{a}}, ArXiv e-prints, arXiv:1502.01586

\bibitem[{{Planck Collaboration} {et~al.}(2015{\natexlab{b}}){Planck
  Collaboration}, {Adam}, {Ade}, {Aghanim}, {Arnaud}, {Ashdown}, {Aumont},
  {Baccigalupi}, {Banday}, {Barreiro}, \& et~al.}]{PlanckCollaboration2015e}
---. 2015{\natexlab{b}}, ArXiv e-prints, arXiv:1502.01587

\bibitem[{{Planck Collaboration} {et~al.}(2015{\natexlab{c}}){Planck
  Collaboration}, {Adam}, {Ade}, {Aghanim}, {Alves}, {Arnaud}, {Ashdown},
  {Aumont}, {Baccigalupi}, {Banday}, \& et~al.}]{PlanckMaps2015}
---. 2015{\natexlab{c}}, ArXiv e-prints, arXiv:1502.01588

\bibitem[{{Planck Collaboration} {et~al.}(2015{\natexlab{d}}){Planck
  Collaboration}, {Ade}, {Aghanim}, {Arnaud}, {Ashdown}, {Aumont},
  {Baccigalupi}, {Banday}, {Barreiro}, {Bartlett}, \&
  et~al.}]{PlanckCollaboration2015a}
{Planck Collaboration}, {Ade}, P.~A.~R., {Aghanim}, N., {et~al.}
  2015{\natexlab{d}}, ArXiv e-prints, arXiv:1502.01589

\bibitem[{{Planck Collaboration} {et~al.}(2015{\natexlab{e}}){Planck
  Collaboration}, {Ade}, {Aghanim}, {Arnaud}, {Ashdown}, {Aumont},
  {Baccigalupi}, {Banday}, {Barreiro}, {Barrena}, \&
  et~al.}]{PlanckCollaboration2015b}
---. 2015{\natexlab{e}}, ArXiv e-prints, arXiv:1502.01598

\bibitem[{{Planck Collaboration} {et~al.}(2015{\natexlab{f}}){Planck
  Collaboration}, {Ade}, {Aghanim}, {Arnaud}, {Ashdown}, {Aumont},
  {Baccigalupi}, {Banday}, {Barreiro}, {Bartolo}, \&
  et~al.}]{PlanckCollaboration2015c}
---. 2015{\natexlab{f}}, ArXiv e-prints, arXiv:1502.01599

\bibitem[{{Planck Collaboration} {et~al.}(2016){Planck Collaboration}, {Ade},
  {Aghanim}, {Arnaud}, {Ashdown}, {Aubourg}, {Aumont}, {Baccigalupi}, {Banday},
  {Barreiro}, {Bartolo}, {Battaner}, {Benabed}, {Benoit-L{\'e}vy},
  {Bersanelli}, {Bielewicz}, {Bock}, {Bonaldi}, {Bonavera}, {Bond}, {Borrill},
  {Bouchet}, {Burigana}, {Calabrese}, {Cardoso}, {Catalano}, {Chamballu},
  {Chiang}, {Christensen}, {Clements}, {Colombo}, {Combet}, {Crill}, {Curto},
  {Cuttaia}, {Danese}, {Davies}, {Davis}, {de Bernardis}, {de Zotti},
  {Delabrouille}, {Dickinson}, {Diego}, {Dolag}, {Donzelli}, {Dor{\'e}},
  {Douspis}, {Ducout}, {Dupac}, {Efstathiou}, {Elsner}, {En{\ss}lin},
  {Eriksen}, {Finelli}, {Forni}, {Frailis}, {Fraisse}, {Franceschi}, {Frejsel},
  {Galeotta}, {Galli}, {Ganga}, {G{\'e}nova-Santos}, {Giard}, {Gjerl{\o}w},
  {Gonz{\'a}lez-Nuevo}, {G{\'o}rski}, {Gregorio}, {Gruppuso}, {Hansen},
  {Harrison}, {Henrot-Versill{\'e}}, {Hern{\'a}ndez-Monteagudo}, {Herranz},
  {Hildebrandt}, {Hivon}, {Hobson}, {Hornstrup}, {Huffenberger}, {Hurier},
  {Jaffe}, {Jaffe}, {Jones}, {Juvela}, {Keih{\"a}nen}, {Keskitalo}, {Kitaura},
  {Kneissl}, {Knoche}, {Kunz}, {Kurki-Suonio}, {Lagache}, {Lamarre}, {Lasenby},
  {Lattanzi}, {Lawrence}, {Leonardi}, {Le{\'o}n-Tavares}, {Levrier}, {Liguori},
  {Lilje}, {Linden-V{\o}rnle}, {L{\'o}pez-Caniego}, {Lubin}, {Ma},
  {Mac{\'{\i}}as-P{\'e}rez}, {Maffei}, {Maino}, {Mak}, {Mandolesi}, {Mangilli},
  {Maris}, {Martin}, {Mart{\'{\i}}nez-Gonz{\'a}lez}, {Masi}, {Matarrese},
  {McGehee}, {Melchiorri}, {Mennella}, {Migliaccio}, {Miville-Desch{\^e}nes},
  {Moneti}, {Montier}, {Morgante}, {Mortlock}, {Munshi}, {Murphy}, {Naselsky},
  {Nati}, {Natoli}, {Noviello}, {Novikov}, {Novikov}, {Oxborrow}, {Pagano},
  {Pajot}, {Paoletti}, {Perdereau}, {Perotto}, {Pettorino}, {Piacentini},
  {Piat}, {Pierpaoli}, {Pointecouteau}, {Polenta}, {Ponthieu}, {Pratt},
  {Puget}, {Puisieux}, {Rachen}, {Racine}, {Reach}, {Reinecke}, {Remazeilles},
  {Renault}, {Renzi}, {Ristorcelli}, {Rocha}, {Rosset}, {Rossetti}, {Roudier},
  {Rubi{\~n}o-Mart{\'{\i}}n}, {Rusholme}, {Sandri}, {Santos}, {Savelainen},
  {Savini}, {Scott}, {Spencer}, {Stolyarov}, {Sudiwala}, {Sunyaev}, {Sutton},
  {Suur-Uski}, {Sygnet}, {Tauber}, {Terenzi}, {Toffolatti}, {Tomasi}, {Tucci},
  {Valenziano}, {Valiviita}, {Van Tent}, {Vielva}, {Villa}, {Wade}, {Wandelt},
  {Wang}, {Wehus}, {Yvon}, {Zacchei}, \& {Zonca}}]{PlanckCollaboration2016}
---. 2016, \aap, 586, A140

\bibitem[{{Platania} {et~al.}(2002){Platania}, {Burigana}, {De Zotti},
  {Lazzaro}, \& {Bersanelli}}]{Platania2002}
{Platania}, P., {Burigana}, C., {De Zotti}, G., {Lazzaro}, E., \& {Bersanelli},
  M. 2002, \mnras, 337, 242

\bibitem[{{Rafferty} {et~al.}(2008){Rafferty}, {McNamara}, \&
  {Nulsen}}]{Rafferty2008}
{Rafferty}, D.~A., {McNamara}, B.~R., \& {Nulsen}, P.~E.~J. 2008, \apj, 687,
  899

\bibitem[{{Rafferty} {et~al.}(2006){Rafferty}, {McNamara}, {Nulsen}, \&
  {Wise}}]{Rafferty2006}
{Rafferty}, D.~A., {McNamara}, B.~R., {Nulsen}, P.~E.~J., \& {Wise}, M.~W.
  2006, \apj, 652, 216

\bibitem[{{Rees} \& {Ostriker}(1977)}]{Rees1977}
{Rees}, M.~J., \& {Ostriker}, J.~P. 1977, \mnras, 179, 541

\bibitem[{{Richstone} {et~al.}(1998){Richstone}, {Ajhar}, {Bender}, {Bower},
  {Dressler}, {Faber}, {Filippenko}, {Gebhardt}, {Green}, {Ho}, {Kormendy},
  {Lauer}, {Magorrian}, \& {Tremaine}}]{Richstone1998}
{Richstone}, D., {Ajhar}, E.~A., {Bender}, R., {et~al.} 1998, \nat, 395, A14

\bibitem[{{Rosas-Guevara} {et~al.}(2016){Rosas-Guevara}, {Bower}, {Schaye},
  {McAlpine}, {Dalla Vecchia}, {Frenk}, {Schaller}, \&
  {Theuns}}]{RosasGuevara2016}
{Rosas-Guevara}, Y., {Bower}, R.~G., {Schaye}, J., {et~al.} 2016, \mnras, 462,
  190

\bibitem[{{Ruan} {et~al.}(2015){Ruan}, {McQuinn}, \& {Anderson}}]{Ruan2015}
{Ruan}, J.~J., {McQuinn}, M., \& {Anderson}, S.~F. 2015, \apj, 802, 135

\bibitem[{{Scannapieco} \& {Oh}(2004)}]{Scannapieco2004}
{Scannapieco}, E., \& {Oh}, S.~P. 2004, \apj, 608, 62

\bibitem[{{Scannapieco} {et~al.}(2005){Scannapieco}, {Silk}, \&
  {Bouwens}}]{Scannapieco2005}
{Scannapieco}, E., {Silk}, J., \& {Bouwens}, R. 2005, \apjl, 635, L13

\bibitem[{{Scannapieco} {et~al.}(2008){Scannapieco}, {Thacker}, \&
  {Couchman}}]{Scannapieco2008}
{Scannapieco}, E., {Thacker}, R.~J., \& {Couchman}, H.~M.~P. 2008, \apj, 678,
  674

\bibitem[{{Schaan} {et~al.}(2016){Schaan}, {Ferraro}, {Vargas-Maga{\~n}a},
  {Smith}, {Ho}, {Aiola}, {Battaglia}, {Bond}, {De Bernardis}, {Calabrese},
  {Cho}, {Devlin}, {Dunkley}, {Gallardo}, {Hasselfield}, {Henderson}, {Hill},
  {Hincks}, {Hlozek}, {Hubmayr}, {Hughes}, {Irwin}, {Koopman}, {Kosowsky},
  {Li}, {Louis}, {Lungu}, {Madhavacheril}, {Maurin}, {McMahon}, {Moodley},
  {Naess}, {Nati}, {Newburgh}, {Niemack}, {Page}, {Pappas}, {Partridge},
  {Schmitt}, {Sehgal}, {Sherwin}, {Sievers}, {Spergel}, {Staggs}, {van
  Engelen}, {Wollack}, \& {ACTPol Collaboration}}]{Schaan2016}
{Schaan}, E., {Ferraro}, S., {Vargas-Maga{\~n}a}, M., {et~al.} 2016, \prd, 93,
  082002

\bibitem[{{Schaffer} {et~al.}(2011){Schaffer}, {Crawford}, {Aird}, {Benson},
  {Bleem}, {Carlstrom}, {Chang}, {Cho}, {Crites}, {de Haan}, {Dobbs}, {George},
  {Halverson}, {Holder}, {Holzapfel}, {Hoover}, {Hrubes}, {Joy}, {Keisler},
  {Knox}, {Lee}, {Leitch}, {Lueker}, {Luong-Van}, {McMahon}, {Mehl}, {Meyer},
  {Mohr}, {Montroy}, {Padin}, {Plagge}, {Pryke}, {Reichardt}, {Ruhl},
  {Shirokoff}, {Spieler}, {Stalder}, {Staniszewski}, {Stark}, {Story},
  {Vanderlinde}, {Vieira}, \& {Williamson}}]{Schaffer2011}
{Schaffer}, K.~K., {Crawford}, T.~M., {Aird}, K.~A., {et~al.} 2011, \apj, 743,
  90

\bibitem[{{Schawinski} {et~al.}(2007){Schawinski}, {Thomas}, {Sarzi},
  {Maraston}, {Kaviraj}, {Joo}, {Yi}, \& {Silk}}]{Schawinski2007}
{Schawinski}, K., {Thomas}, D., {Sarzi}, M., {et~al.} 2007, \mnras, 382, 1415

\bibitem[{{Schaye} {et~al.}(2015){Schaye}, {Crain}, {Bower}, {Furlong},
  {Schaller}, {Theuns}, {Dalla Vecchia}, {Frenk}, {McCarthy}, {Helly},
  {Jenkins}, {Rosas-Guevara}, {White}, {Baes}, {Booth}, {Camps}, {Navarro},
  {Qu}, {Rahmati}, {Sawala}, {Thomas}, \& {Trayford}}]{Schaye2015}
{Schaye}, J., {Crain}, R.~A., {Bower}, R.~G., {et~al.} 2015, \mnras, 446, 521

\bibitem[{{Schlegel} {et~al.}(2016){Schlegel}, {Jones}, {Machacek}, \&
  {Vega}}]{Schlegel2016}
{Schlegel}, E.~M., {Jones}, C., {Machacek}, M., \& {Vega}, L.~D. 2016, \apj,
  823, 75

\bibitem[{{Sijacki} {et~al.}(2007){Sijacki}, {Springel}, {Di Matteo}, \&
  {Hernquist}}]{Sijacki2007}
{Sijacki}, D., {Springel}, V., {Di Matteo}, T., \& {Hernquist}, L. 2007,
  \mnras, 380, 877

\bibitem[{{Silk}(1968)}]{Silk1968}
{Silk}, J. 1968, \apj, 151, 459

\bibitem[{{Simionescu} {et~al.}(2009){Simionescu}, {Roediger}, {Nulsen},
  {Br{\"u}ggen}, {Forman}, {B{\"o}hringer}, {Werner}, \&
  {Finoguenov}}]{Simionescu2009}
{Simionescu}, A., {Roediger}, E., {Nulsen}, P.~E.~J., {et~al.} 2009, \aap, 495,
  721

\bibitem[{{Siudek} {et~al.}(2016){Siudek}, {Ma{\l}ek}, {Scodeggio}, {Garilli},
  {Pollo}, {Haines}, {Fritz}, {Bolzonella}, {de la Torre}, {Granett}, {Guzzo},
  {Abbas}, {Adami}, {Bottini}, {Cappi}, {Cucciati}, {De Lucia}, {Davidzon},
  {Franzetti}, {Iovino}, {Krywult}, {Le Brun}, {Le F{\`e}vre}, {Maccagni},
  {Marchetti}, {Marulli}, {Polletta}, {Tasca}, {Tojeiro}, {Vergani},
  {Zanichelli}, {Arnouts}, {Bel}, {Branchini}, {Ilbert}, {Gargiulo},
  {Moscardini}, {Takeuchi}, \& {Zamorani}}]{Siudek2016}
{Siudek}, M., {Ma{\l}ek}, K., {Scodeggio}, M., {et~al.} 2016, ArXiv e-prints,
  arXiv:1605.05503

\bibitem[{{Skelton} {et~al.}(2014){Skelton}, {Whitaker}, {Momcheva}, {Brammer},
  {van Dokkum}, {Labb{\'e}}, {Franx}, {van der Wel}, {Bezanson}, {Da Cunha},
  {Fumagalli}, {F{\"o}rster Schreiber}, {Kriek}, {Leja}, {Lundgren}, {Magee},
  {Marchesini}, {Maseda}, {Nelson}, {Oesch}, {Pacifici}, {Patel}, {Price},
  {Rix}, {Tal}, {Wake}, \& {Wuyts}}]{Skelton2014}
{Skelton}, R.~E., {Whitaker}, K.~E., {Momcheva}, I.~G., {et~al.} 2014, \apjs,
  214, 24

\bibitem[{{Soergel} {et~al.}(2016){Soergel}, {Flender}, {Story}, {Bleem},
  {Giannantonio}, {Efstathiou}, {Rykoff}, {Benson}, {Crawford}, {Dodelson},
  {Habib}, {Heitmann}, {Holder}, {Jain}, {Rozo}, {Saro}, {Weller}, {Abdalla},
  {Allam}, {Annis}, {Armstrong}, {Benoit-L{\'e}vy}, {Bernstein}, {Carlstrom},
  {Carnero Rosell}, {Carrasco Kind}, {Castander}, {Chiu}, {Chown}, {Crocce},
  {Cunha}, {D'Andrea}, {da Costa}, {de Haan}, {Desai}, {Diehl}, {Dietrich},
  {Doel}, {Estrada}, {Evrard}, {Flaugher}, {Fosalba}, {Frieman}, {Gaztanaga},
  {Gruen}, {Gruendl}, {Holzapfel}, {Honscheid}, {James}, {Keisler}, {Kuehn},
  {Kuropatkin}, {Lahav}, {Lima}, {Marshall}, {McDonald}, {Melchior}, {Miller},
  {Miquel}, {Nord}, {Ogando}, {Omori}, {Plazas}, {Rapetti}, {Reichardt},
  {Romer}, {Roodman}, {Saliwanchik}, {Sanchez}, {Schubnell}, {Sevilla-Noarbe},
  {Sheldon}, {Smith}, {Soares-Santos}, {Sobreira}, {Stark}, {Suchyta},
  {Swanson}, {Tarle}, {Thomas}, {Vieira}, {Walker}, {Whitehorn}, {DES
  Collaboration}, \& {SPT Collaboration}}]{Soergel2016}
{Soergel}, B., {Flender}, S., {Story}, K.~T., {et~al.} 2016, \mnras, 461, 3172

\bibitem[{{Spacek} {et~al.}(2016){Spacek}, {Scannapieco}, {Cohen}, {Joshi}, \&
  {Mauskopf}}]{Spacek2016}
{Spacek}, A., {Scannapieco}, E., {Cohen}, S., {Joshi}, B., \& {Mauskopf}, P.
  2016, \apj, 819, 128

\bibitem[{{Sturm} {et~al.}(2011){Sturm}, {Gonz{\'a}lez-Alfonso}, {Veilleux},
  {Fischer}, {Graci{\'a}-Carpio}, {Hailey-Dunsheath}, {Contursi}, {Poglitsch},
  {Sternberg}, {Davies}, {Genzel}, {Lutz}, {Tacconi}, {Verma}, {Maiolino}, \&
  {de Jong}}]{Sturm2011}
{Sturm}, E., {Gonz{\'a}lez-Alfonso}, E., {Veilleux}, S., {et~al.} 2011, \apjl,
  733, L16

\bibitem[{{Sunyaev} \& {Zeldovich}(1970)}]{Sunyaev1970}
{Sunyaev}, R.~A., \& {Zeldovich}, Y.~B. 1970, \apss, 7, 3

\bibitem[{{Sunyaev} \& {Zeldovich}(1972)}]{Sunyaev1972}
---. 1972, Comments on Astrophysics and Space Physics, 4, 173

\bibitem[{{Swetz} {et~al.}(2011){Swetz}, {Ade}, {Amiri}, {Appel},
  {Battistelli}, {Burger}, {Chervenak}, {Devlin}, {Dicker}, {Doriese},
  {D{\"u}nner}, {Essinger-Hileman}, {Fisher}, {Fowler}, {Halpern},
  {Hasselfield}, {Hilton}, {Hincks}, {Irwin}, {Jarosik}, {Kaul}, {Klein},
  {Lau}, {Limon}, {Marriage}, {Marsden}, {Martocci}, {Mauskopf}, {Moseley},
  {Netterfield}, {Niemack}, {Nolta}, {Page}, {Parker}, {Staggs}, {Stryzak},
  {Switzer}, {Thornton}, {Tucker}, {Wollack}, \& {Zhao}}]{Swetz2011}
{Swetz}, D.~S., {Ade}, P.~A.~R., {Amiri}, M., {et~al.} 2011, \apjs, 194, 41

\bibitem[{{Teimoorinia} {et~al.}(2016){Teimoorinia}, {Bluck}, \&
  {Ellison}}]{Teimoorinia2016}
{Teimoorinia}, H., {Bluck}, A.~F.~L., \& {Ellison}, S.~L. 2016, \mnras, 457,
  2086

\bibitem[{{Thacker} {et~al.}(2006){Thacker}, {Scannapieco}, \&
  {Couchman}}]{Thacker2006}
{Thacker}, R.~J., {Scannapieco}, E., \& {Couchman}, H.~M.~P. 2006, \apj, 653,
  86

\bibitem[{{Tombesi} {et~al.}(2015){Tombesi}, {Mel{\'e}ndez}, {Veilleux},
  {Reeves}, {Gonz{\'a}lez-Alfonso}, \& {Reynolds}}]{Tombesi2015}
{Tombesi}, F., {Mel{\'e}ndez}, M., {Veilleux}, S., {et~al.} 2015, \nat, 519,
  436

\bibitem[{{Treu} {et~al.}(2005){Treu}, {Ellis}, {Liao}, \& {van
  Dokkum}}]{Treu2005}
{Treu}, T., {Ellis}, R.~S., {Liao}, T.~X., \& {van Dokkum}, P.~G. 2005, \apjl,
  622, L5

\bibitem[{{Ueda} {et~al.}(2003){Ueda}, {Akiyama}, {Ohta}, \&
  {Miyaji}}]{Ueda2003}
{Ueda}, Y., {Akiyama}, M., {Ohta}, K., \& {Miyaji}, T. 2003, \apj, 598, 886

\bibitem[{{van der Wel} {et~al.}(2012){van der Wel}, {Bell}, {H{\"a}ussler},
  {McGrath}, {Chang}, {Guo}, {McIntosh}, {Rix}, {Barden}, {Cheung}, {Faber},
  {Ferguson}, {Galametz}, {Grogin}, {Hartley}, {Kartaltepe}, {Kocevski},
  {Koekemoer}, {Lotz}, {Mozena}, {Peth}, \& {Peng}}]{vanderWel2012}
{van der Wel}, A., {Bell}, E.~F., {H{\"a}ussler}, B., {et~al.} 2012, \apjs,
  203, 24

\bibitem[{{Veilleux} {et~al.}(2013){Veilleux}, {Mel{\'e}ndez}, {Sturm},
  {Gracia-Carpio}, {Fischer}, {Gonz{\'a}lez-Alfonso}, {Contursi}, {Lutz},
  {Poglitsch}, {Davies}, {Genzel}, {Tacconi}, {de Jong}, {Sternberg}, {Netzer},
  {Hailey-Dunsheath}, {Verma}, {Rupke}, {Maiolino}, {Teng}, \&
  {Polisensky}}]{Veilleux2013}
{Veilleux}, S., {Mel{\'e}ndez}, M., {Sturm}, E., {et~al.} 2013, \apj, 776, 27

\bibitem[{{Vergani} {et~al.}(2008){Vergani}, {Scodeggio}, {Pozzetti}, {Iovino},
  {Franzetti}, {Garilli}, {Zamorani}, {Maccagni}, {Lamareille}, {Le F{\`e}vre},
  {Charlot}, {Contini}, {Guzzo}, {Bottini}, {Le Brun}, {Picat}, {Scaramella},
  {Tresse}, {Vettolani}, {Zanichelli}, {Adami}, {Arnouts}, {Bardelli},
  {Bolzonella}, {Cappi}, {Ciliegi}, {Foucaud}, {Gavignaud}, {Ilbert},
  {McCracken}, {Marano}, {Marinoni}, {Mazure}, {Meneux}, {Merighi}, {Paltani},
  {Pell{\`o}}, {Pollo}, {Radovich}, {Zucca}, {Bondi}, {Bongiorno},
  {Brinchmann}, {Cucciati}, {de la Torre}, {Gregorini}, {Perez-Montero},
  {Mellier}, {Merluzzi}, \& {Temporin}}]{Vergani2008}
{Vergani}, D., {Scodeggio}, M., {Pozzetti}, L., {et~al.} 2008, \aap, 487, 89

\bibitem[{{Voges} {et~al.}(1999){Voges}, {Aschenbach}, {Boller},
  {Br{\"a}uninger}, {Briel}, {Burkert}, {Dennerl}, {Englhauser}, {Gruber},
  {Haberl}, {Hartner}, {Hasinger}, {K{\"u}rster}, {Pfeffermann}, {Pietsch},
  {Predehl}, {Rosso}, {Schmitt}, {Tr{\"u}mper}, \& {Zimmermann}}]{Voges1999}
{Voges}, W., {Aschenbach}, B., {Boller}, T., {et~al.} 1999, \aap, 349, 389

\bibitem[{{Voit}(1994)}]{Voit1994}
{Voit}, G.~M. 1994, \apjl, 432, L19

\bibitem[{{Wampler} {et~al.}(1995){Wampler}, {Chugai}, \&
  {Petitjean}}]{Wampler1995}
{Wampler}, E.~J., {Chugai}, N.~N., \& {Petitjean}, P. 1995, \apj, 443, 586

\bibitem[{{White} \& {Frenk}(1991)}]{White1991}
{White}, S.~D.~M., \& {Frenk}, C.~S. 1991, \apj, 379, 52

\bibitem[{{Wright} {et~al.}(2010){Wright}, {Eisenhardt}, {Mainzer}, {Ressler},
  {Cutri}, {Jarrett}, {Kirkpatrick}, {Padgett}, {McMillan}, {Skrutskie},
  {Stanford}, {Cohen}, {Walker}, {Mather}, {Leisawitz}, {Gautier}, {McLean},
  {Benford}, {Lonsdale}, {Blain}, {Mendez}, {Irace}, {Duval}, {Liu}, {Royer},
  {Heinrichsen}, {Howard}, {Shannon}, {Kendall}, {Walsh}, {Larsen}, {Cardon},
  {Schick}, {Schwalm}, {Abid}, {Fabinsky}, {Naes}, \& {Tsai}}]{Wright2010}
{Wright}, E.~L., {Eisenhardt}, P.~R.~M., {Mainzer}, A.~K., {et~al.} 2010, \aj,
  140, 1868

\bibitem[{{Yamamura} {et~al.}(2010){Yamamura}, {Makiuti}, {Ikeda}, {Fukuda},
  {Oyabu}, {Koga}, \& {White}}]{Yamamura2010}
{Yamamura}, I., {Makiuti}, S., {Ikeda}, N., {et~al.} 2010, VizieR Online Data
  Catalog, 2298

\end{thebibliography}
\end{document}